\documentclass[twocolumn]{aastex631}

\shorttitle{Strongly lensed quasars in KiDS DR5}
\shortauthors{Z.He et al.}
\graphicspath{{./}{figures/}}

\begin{document}

\title{Using Convolutional Neural Networks to Search for Strongly Lensed Quasars in KiDS DR5}

\email{zzhe@pmo.ac.cn}
\email{liruiww@gmail.com}

\author[0000-0001-8554-9163]{Zizhao He}
\affiliation{Purple Mountain Observatory, Chinese Academy of Sciences, Nanjing, Jiangsu, 210023, China}

\author[0000-0002-3490-4089]{Rui Li}
\affiliation{Institute for Astrophysics, School of Physics, Zhengzhou University, Zhengzhou, 450001, China}

\author[0000-0002-9063-698X]{Yiping Shu}
\affiliation{Purple Mountain Observatory, Chinese Academy of Sciences, Nanjing, Jiangsu, 210023, China}

\author[0000-0001-7958-6531]{Crescenzo Tortora}
\affiliation{INAF -- Osservatorio Astronomico di Capodimonte, Salita Moiariello 16, I-80131, Napoli, Italy}

\author[0000-0002-8700-3671]{Xinzhong Er}
\affiliation{Tianjin Astrophysics Center, Tianjin Normal University, Tianjin 300387, P.R.China}

\author[0000-0002-2468-5169]{Raoul Ca\~nameras}
\affiliation{Max-Planck-Institut f{\"u}r Astrophysik, Karl-Schwarzschild Stra{\ss}e 1, 85748 Garching, Germany}
\affiliation{Technical University of Munich, TUM School of Natural Sciences, Department of Physics,  James-Franck-Stra{\ss}e 1, 85748 Garching, Germany}
\affiliation{Aix Marseille Univ, CNRS, CNES, LAM, Marseille, France}

\author[0000-0003-2497-6334]{Stefan Schuldt}
\affiliation{Dipartimento di Fisica, Universit\`a  degli Studi di Milano, via Celoria 16, I-20133 Milano, Italy}
\affiliation{INAF - IASF Milano, via A. Corti 12, I-20133 Milano, Italy}

\author[0000-0003-0911-8884]{Nicola R. Napolitano}
\affiliation{School of Physics and Astronomy, Sun Yat-sen University, Zhuhai Campus, 2 Daxue Road, Xiangzhou District, Zhuhai, People's Republic of China}
\affiliation{CSST Science Center for Guangdong-Hong Kong-Macau Great Bay Area, Zhuhai, 519082, People's Republic of China}
\affiliation{INAF -- Osservatorio Astronomico di Capodimonte, Salita Moiariello 16, 80131 - Napoli, Italy}

\author{Bharath Chowdhary N}
\affiliation{Kapteyn Astronomical Institute, University of Groningen, PO Box 800, NL-9700 AV Groningen, the Netherlands}

\author[0009-0006-9345-9639]{Qihang Chen}
\affiliation{Department of Astronomy, Beijing Normal University, Beijing 100875, China}
\affiliation{Institute for Frontier in Astronomy and Astrophysics, Beijing Normal University, Beijing, 102206, China}

\author{Nan Li}
\affiliation{Key Laboratory of Space Astronomy and Technology, National Astronomical Observatories, CAS, Beijing 100101, China}

\affiliation{INAF - IASF Milano, via A. Corti 12, I-20133 Milano, Italy}

\author[0000-0002-1530-2680]{Haicheng Feng}
\affiliation{Yunnan Observatories, Chinese Academy of Sciences, Kunming 650216, Yunnan, People's Republic of China}

\author{Limeng Deng}
\affiliation{Purple Mountain Observatory, Chinese Academy of Sciences, Nanjing, Jiangsu, 210023, China}
\affiliation{School of Astronomy and Space Sciences, University of Science and Technology of China, Hefei 230026}

\author{Guoliang Li}
\affiliation{Purple Mountain Observatory, Chinese Academy of Sciences, Nanjing, Jiangsu, 210023, China}

\author{L.V.E. Koopmans}
\affiliation{Kapteyn Astronomical Institute, University of Groningen, PO Box 800, NL-9700 AV Groningen, the Netherlands}

\author{Andrej Dvornik}
\affiliation{Ruhr University Bochum, Faculty of Physics and Astronomy,Astronomical Institute (AIRUB), German Centre for Cosmological Lensing, 44780 Bochum, Germany}



\begin{abstract}

Gravitationally strongly lensed quasars (SL-QSO) offer invaluable insights into cosmological and astrophysical phenomena. With the data from ongoing and next-generation surveys, thousands of SL-QSO systems can be discovered expectedly, leading to unprecedented opportunities. However, the challenge lies in identifying SL-QSO from enormous datasets with high recall and purity in an automated and efficient manner. Hence, we developed a program based on a Convolutional Neural Network (CNN) for finding SL-QSO from large-scale surveys and applied it to the Kilo-degree Survey Data Release 5 (KiDS DR5). Our approach involves three key stages: firstly, we pre-selected ten million bright objects (with $r$-band $\tt{MAG\_AUTO} < 22$), excluding stars from the dataset; secondly, we established realistic training and test sets to train and fine-tune the CNN, resulting in the identification of 4195 machine candidates, and the false positive rate (FPR) of $\sim$1/2000 and recall of 0.8125 evaluated by using the real test set containing 16 confirmed lensed quasars; thirdly, human inspections were performed for further selections, and then 272 SL-QSO candidates were eventually found in total, including 16 high-score, 118 median-score, and 138 lower-score candidates, separately. Removing the systems already confirmed or identified in other papers, we end up with 229 SL-QSO candidates, including 7 high-score, 95 median-score, and 127 lower-score candidates, and the corresponding catalog is publicly available online.\footnote{https://github.com/EigenHermit/H24} We have also included an excellent quad candidate in the appendix, discovered serendipitously during the fine-tuning process of the CNN.



\end{abstract}



\section{Introduction} \label{sec:intro}

Strongly lensed quasars serve as a crucial cosmological tool for investigating a wide range of
astrophysical and cosmological phenomena. For instance, by studying the microlensing effect caused by stars in the lensing galaxies, one can uncover valuable information about the accretion disk properties of the background Active Galactic Nuclei \citep{Anguita2008, Sluse2011, Motta2012, Guerras2013, Braibant2014, Fian2021, Hutsemekers2021}. Lens modeling techniques allow researchers to constrain the distribution of luminous and dark matter within the lens galaxies \citep{Oguri2014, Suyu2014, Sonnenfeld2021, Vyvere2021}. Additionally, by incorporating time-domain data, one can measure the time delay between the multiple images, which provides a direct measurement of the Hubble-Lemaître constant \citep[$H_0$,][]{Suyu2014, Liao2019, Suyu2017, Zhu2023}. This application is particularly important in addressing the recent Hubble tension \citep{Riess2019, Verde2019, Shah2021}, which refers to the discrepancy in $H_0$ values derived from high and low redshift observations. Strong lensing time delay offers an independent measurement that is not reliant on the cosmic microwave background \citep[CMB,][]{Bennett2013, Planck_Collaboration2020} or the local distance ladder \citep[see, e.g.,][]{Riess2016, Riess2018, Riess2019, Yuan2019, Freedman2019}.

However, the limited number of known lensed quasars greatly restricts the ability to investigate these aforementioned topics. For instance, to achieve a one-percent precision in constraining $H_0$,  it is suggested that at least 40 quadruply-imaged lensed quasars with wide separations and high-precision time-delay measurements are needed \citep{Shajib2020}. However, such systems are rare, with only around seven meeting these specific criteria out of the existing $\sim$300 known lensed quasars.

The search for lensed quasars in optical imaging surveys typically involves a two-step process: candidate identification followed by spectroscopic confirmation. Photometric surveys that currently exist or are planned for the future play a pivotal role in identifying a significant number of potential candidates \citep{Chan2015, Spiniello2018, Spiniello2019, Dawes2023, He2023, Andika2023, Chan2023}, which can then undergo further scrutiny for verification \citep[e.g.,][]{Lemon2022}. Prominent surveys such as the Kilo-Degree Survey (KiDS) offer extensive sky coverage, good image quality, and depth, rendering the discovery of numerous new lensed systems feasible \citep{Petrillo2017, Petrillo2019, Li2020, Li2021, Napolitano2020}.

A dozen lensed quasar systems have been identified in KiDS in previous efforts \citep{Spiniello2018, Spiniello2019, Khramtsov2019}. However, the number is significantly below the theoretical prediction \citep{Oguri2010}. Moreover, with the forthcoming release of KiDS DR5, it is anticipated that more systems await detection. To enhance the process of candidate identification, we implemented a machine learning technique known as Convolutional Neural Network (CNN). Our methodology involves training the CNN using a blend of simulated and real data.

Notably, within this field, numerous studies have trained CNNs specifically for detecting galaxy-galaxy strong lenses. These endeavors have led to the discovery of thousands of lens candidates \citep[e.g.,][]{Sonnenfeld2018, Petrillo2019, He2020, Canameras2020, Huang2021, Li2021, Rojas2022}, subsequently validated through investigations \citep[e.g.,][]{Tran2022, Cikota2023, Dux2024}. In fact, in the Strong Lensing Finding Challenge \citep{Metcalf2019}, CNN-based methods outperformed other techniques. Moreover, this approach has been extended to the search for lensed quasars \citep[e.g.,][]{Andika2023,Dux2023}, albeit with fewer studies compared to galaxy-galaxy lens detection.

In this study, we further advanced this methodology by developing a CNN as a lensed quasar-finder that can achieve a very low false positive rate (FPR) while maintaining a high true positive rate (TPR), aimed at detecting candidates within the KiDS DR5 dataset.

The paper is structured as follows. Section \ref{sec:data} delineates the datasets utilized or compiled for this study, encompassing both the data analyzed for identifying lensed quasar candidates and the sets employed for training and testing purposes. Section \ref{sec:method} elaborates on the methodologies employed, including the Network architecture (hereafter, we use the capitalized ``Network" for our specific network) and the training process of the Network. Subsequently, Section \ref{sec:result} presents the human inspection and the obtained results, accompanied by a detailed catalog. The selection effect of the Network of several lensing parameters is delivered in Section \ref{sec:sel_eff}. Finally, the summary and future plans are provided in Section \ref{sec:sum}. In this paper, a fiducial cosmological model with $\Omega_m=0.26$, $\Omega_{DE}=0.74$, $h=0.72$, $w_0=-1$, and $w_a=0$ is assumed. Unless otherwise stated, all magnitudes quoted in this paper are in the AB system.

\section{Data-sets}
\label{sec:data}
Here, we present three datasets compiled for our study. Firstly, to establish a sample conducive to identifying lensed quasar candidates, we compiled the target dataset, consisting of 9,958,196 objects (Section  \ref{sec:tobeapplied_data}). Secondly, we introduce our training and validation set, comprising a total of 26,005 positives and 27,275 negatives (Section \ref{sec:training set}). Of this, 60\% was used for training and 10\% for validation. Lastly, to evaluate the Network's performance, we utilized two test sets (Section \ref{sec: test set}). The simulated test set contains the remaining 30\% of the 26,005 positives and 27,275 negatives and was used to test the Network's selection effect. The other, a real test set, comprises 16×6 lensed quasars and 50,000 non-lensed quasars, where `×6' denotes the data augmentation applied to the lensed quasar samples. Notably, we used $gri$ 3-band fits data as the input data for the Network.
\subsection{Target dataset}
\label{sec:tobeapplied_data}

The Kilo-Degree Survey \citep[KiDS,][]{deJong2013} is a publicly available survey conducted by the European Southern Observatory (ESO). It utilizes five optical broad-band filters: $u$, $g$, $r$, $i_1$, and $i_2$ using the VLT Survey Telescope \citep[VST,][]{Capaccioli2011}. The near-infrared (NIR) data from KiDS is supplemented by data from the VISTA Kilo-Degree Infrared Galaxy Survey \citep[VIKING,][]{Edge2013}, which has completed observations in five NIR bands ($Z$, $Y$, $J$, $H$, and $K_s$) within the same sky region. The last data release, KiDS DR5 \citep{Wright2024}, includes all the survey tiles, totaling 1347 deg$^2$. The typical limiting magnitudes in each band ($u$, $g$, $r$, $i_1$, $i_2$, $Z$, $Y$, $J$, $H$, and $K_s$) are as follows: 24.17, 24.96, 24.79, 23.41, 23.49, 23.47, 22.72, 22.53, 21.94, and 21.75 \citep[5$\sigma$ in $2\arcsec$ aperture,][]{Wright2024}.

We defined the dataset for analysis based on the KiDS DR5 catalogs and images \citep{Wright2024}, incorporating the star-quasar-galaxy classification from \cite{Feng2024}. In \cite{Feng2024}, probabilities $p_{Star}$, $p_{Galaxy}$, and $p_{QSO}$ represent the likelihood of a source being a star, galaxy, or quasar, respectively, derived using machine learning techniques that leverage both morphological and photometric information. This method achieved an impressive overall classification accuracy of 98.76\% for objects with $r$-band $\tt{MAG\_AUTO} < 22$.

The catalog from \cite{Wright2024} was subjected to the following selection criteria. First, we considered only sources with an $r$-band $\tt{MAG\_AUTO} < 22$, as the star-quasar classification in \cite{Feng2024} has higher uncertainty beyond this threshold. A similar cutoff was employed in \cite{Khramtsov2019}. Second, we selected sources where $p_{Star}$ was the lowest of the three probabilities. Specifically, for stars, the method demonstrated a purity of 98.5\% and a completeness of 98.62\%, confirming its effectiveness in accurately identifying and excluding stars. Additionally, our analysis was limited to sources with detections in all three bands: $g$, $r$, and $i$. After applying these selection criteria, we identified 9,958,196 relevant sources out of an initial 65,907,702.

In parallel to the aforementioned target dataset, we identified 191,321 Luminous Red Galaxies (LRGs) out of the 65,907,702 total sources, using the selection criteria outlined in equations (1) and (2) in \cite{Li2021}. This selection methodology aligns with that used in \cite{Petrillo2017, He2020, Li2020}. Many studies that focus on searching for lensed galaxies use LRGs as their primary target dataset, but this is not suitable for our purposes. The reason is that, according to images of spectroscopically confirmed lensed quasars, the lensing galaxies are invisible before the subtraction of quasar images in many systems (such as J0907+0003 and J1037+0018 in Figure \ref{fig:real_lensed quasars}). These systems will not be selected when performing LRG selection. Nevertheless, a significant majority of the LRGs (191,101 out of 191,321) were already included within our target dataset.

\subsection{Training and validation sets}
\label{sec:training set}

Our previous experience \citep[e.g.,][]{He2020,Li2021,Canameras2023}, as well as results from other researchers \citep[e.g.,][]{Huang2021}, indicates that the training set significantly influences the performance of the classifier. Thus, we concentrated on creating a realistic training set. 

Furthermore, we decided to generate training sets only with very obvious strong lensing features under KiDS observational conditions. Because of the seeing, we cannot identify systems with small separations or ambiguous multiple images. Consequently, some lensed quasars were inevitably missed under the current methodology and observational conditions. We anticipate using next-generation datasets such as the China Space Station Telescope \citep[CSST;][]{csst_zhanhu}, Euclid \citep{Euclid-intro, Euclid2024}, and Roman \citep{Roman-intro} to enhance observational capacity for those lensed quasars missed at this stage.

More specifically, we generated the positive part of the training set by combining real LRGs (see Section \ref{sec:tobeapplied_data}) and mock lensed multiple images \citep{Petrillo2017,Li2020,Li2021,He2020,Nagam2023,Andika2023}. We selected 10,000 LRGs from a sample of 191,321 LRGs through visual inspection, choosing those with typical LRG morphology and without lensing features at random. We then estimated their velocity dispersion ($\sigma_v$) using the fundamental plane of elliptical galaxies.

\begin{equation}\label{equ:fp}
    {\rm log_{10}}R_e = a_r{\rm log_{10}}\sigma_v + b_r\mu_r + c_r,
\end{equation}
the index $r$ indicates that we used FP in $r$ band and $a_r$ = 1.207, $b_r$ = 0.315, $c_r$ = -8.463 \citep{Yoon2020}. Furthermore, $\mu_r$ is given by:
\begin{equation}
    \mu_r = m_r+5{\rm log_{10}}r_{e,r}+2.5{\rm log_{10}}2\pi-2.5{\rm log_{10}}(1+z_d)^3.
\end{equation}
where $R_e$ is the physical half-light radius in the unit of kpc, while $r_e$ is in arcsec. In this study, the redshifts of lensing LRG and background quasars were obtained by GaZNet \citep{Li2022GaZNet}. It is a CNN which determine the redshift by using both morphological and photometric information. Similarly, $r_e$ was obtained using the methodologies described in \cite{Li2022GaLNet}, which is a CNN that can determine the parameters of the Sersic profile.

After obtaining the velocity dispersion, we employed the Singular Isothermal Ellipsoid \citep[SIE,][]{Kormann1994} profile to generate multiple images. The convergence of SIE is given by:

\begin{equation}
\kappa(x,y)=\frac{\theta_E}{2}
\frac{1}{\sqrt{qx^2 +y^2/q}},
\end{equation}
where $q$ is the axial ratio between the minor and major axes, which is assumed to be the same in the mock of the light distribution of lensing galaxy and its host halo; $x,y$ are the coordinates in the lens plane. We assume the $\sigma_v$ estimated by Equation \ref{equ:fp} is a good estimation of $\sigma_v$ of SIE halo, this is supported by previous works such as \cite{Bolton2008}. Moreover, $\theta_E$ is the Einstein radius:
\begin{equation}
    \theta_E=4\pi \left(\frac{\sigma_v}{c}\right)^2 \frac{D_{\mathrm{ds}}}{D_{\mathrm{s}}},
    \label{eq:rein}
\end{equation}
where $D_{\mathrm{ds}}$ is the angular diameter distance between lens and source while $D_{\mathrm{d}}$ is the angular diameter distance between lens and the observer. 

Utilizing these models and placing the point-like source randomly in the source plane, we can calculate the positions and magnifications of the multiple images ($\mu_i$, where $i$ represents the $i$-th image). As background sources, we selected samples with a $p_{QSO}$ value greater than 0.9. There are 499,027 objects that satisfy this condition. The fluxes of the multiple images in $gri$ bands are given by $\mu_i f_{i}$, where $f_i$ denotes the un-lensed fluxes of the background sources in the simulated band. We used the methodologies outlined in \cite{Li2022GaZNet} to derive $z_{GaZ}$ of quasars, which are used to represent $z_s$.

\begin{figure*}[htbp]
    \centering
    \includegraphics[scale=0.35]{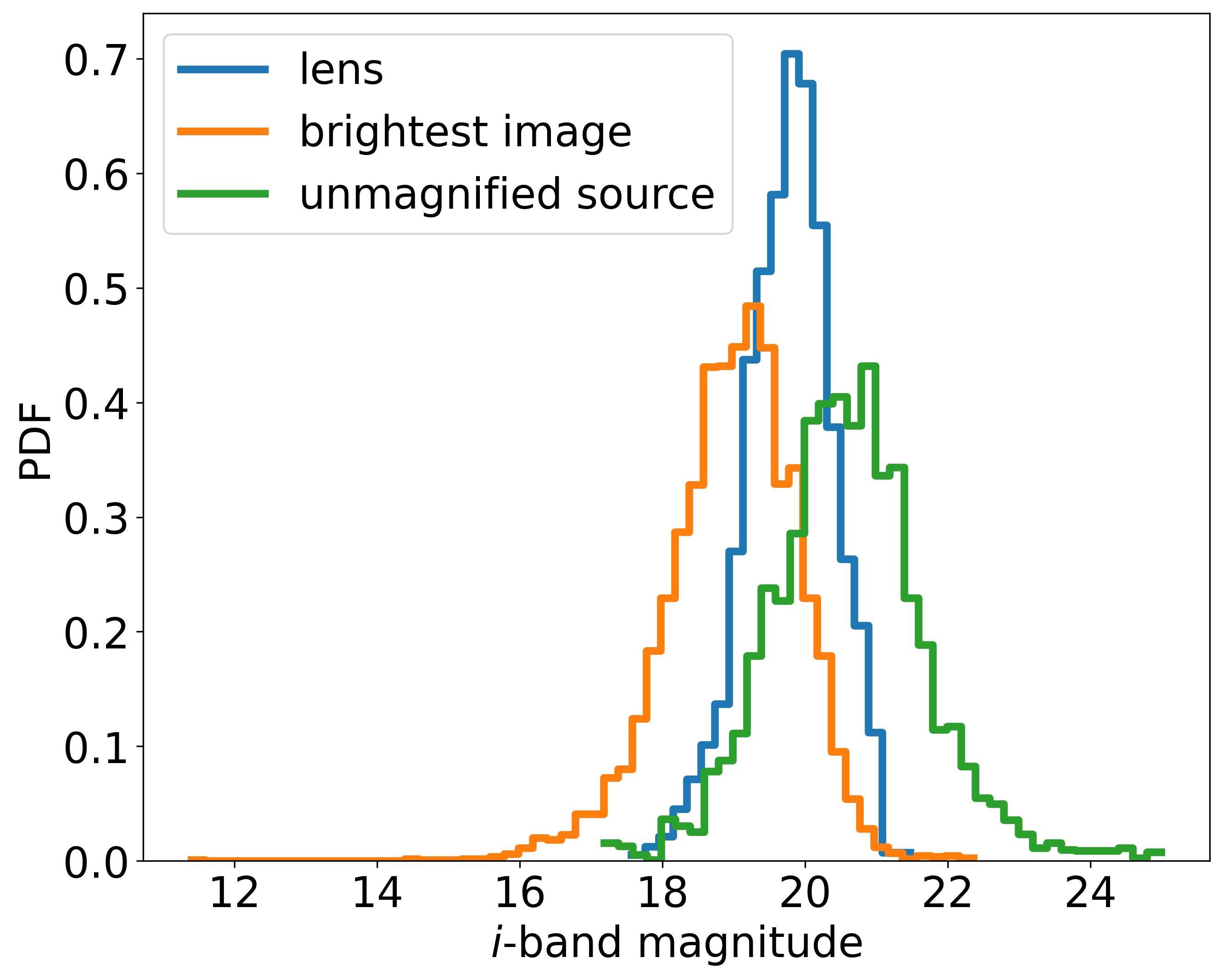}
    \includegraphics[scale=0.35]{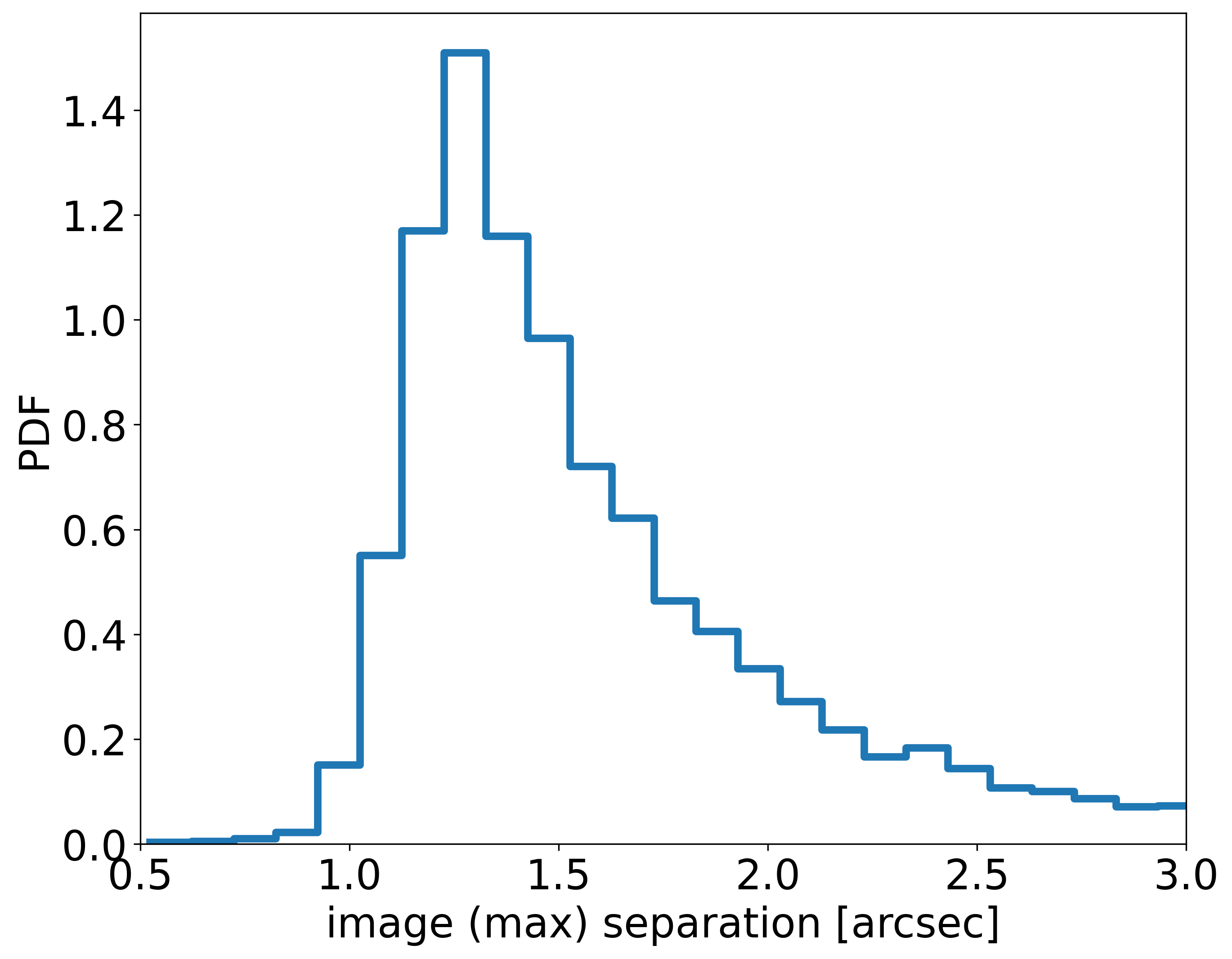}
    \includegraphics[scale=0.35]{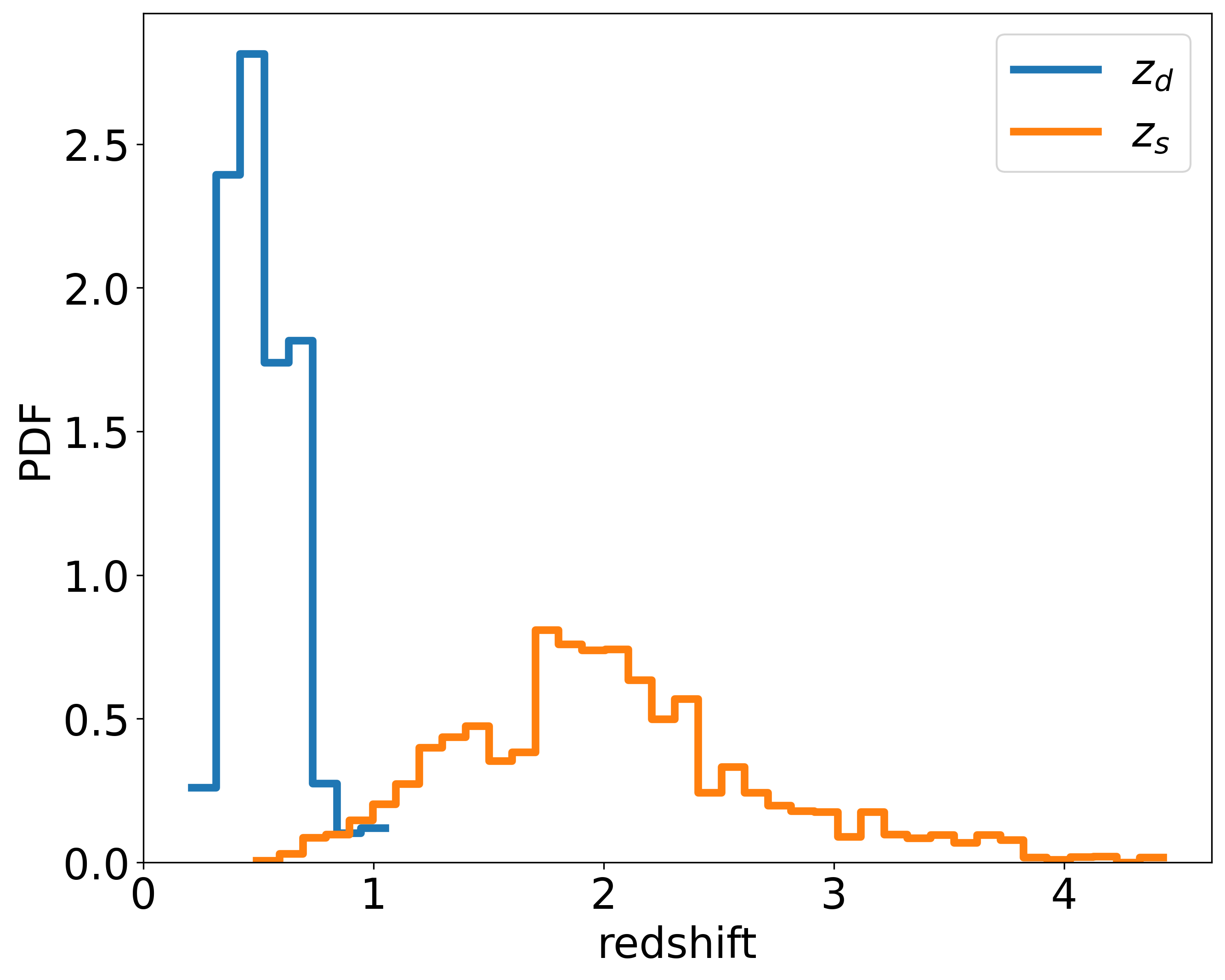}
    \includegraphics[scale=0.35]{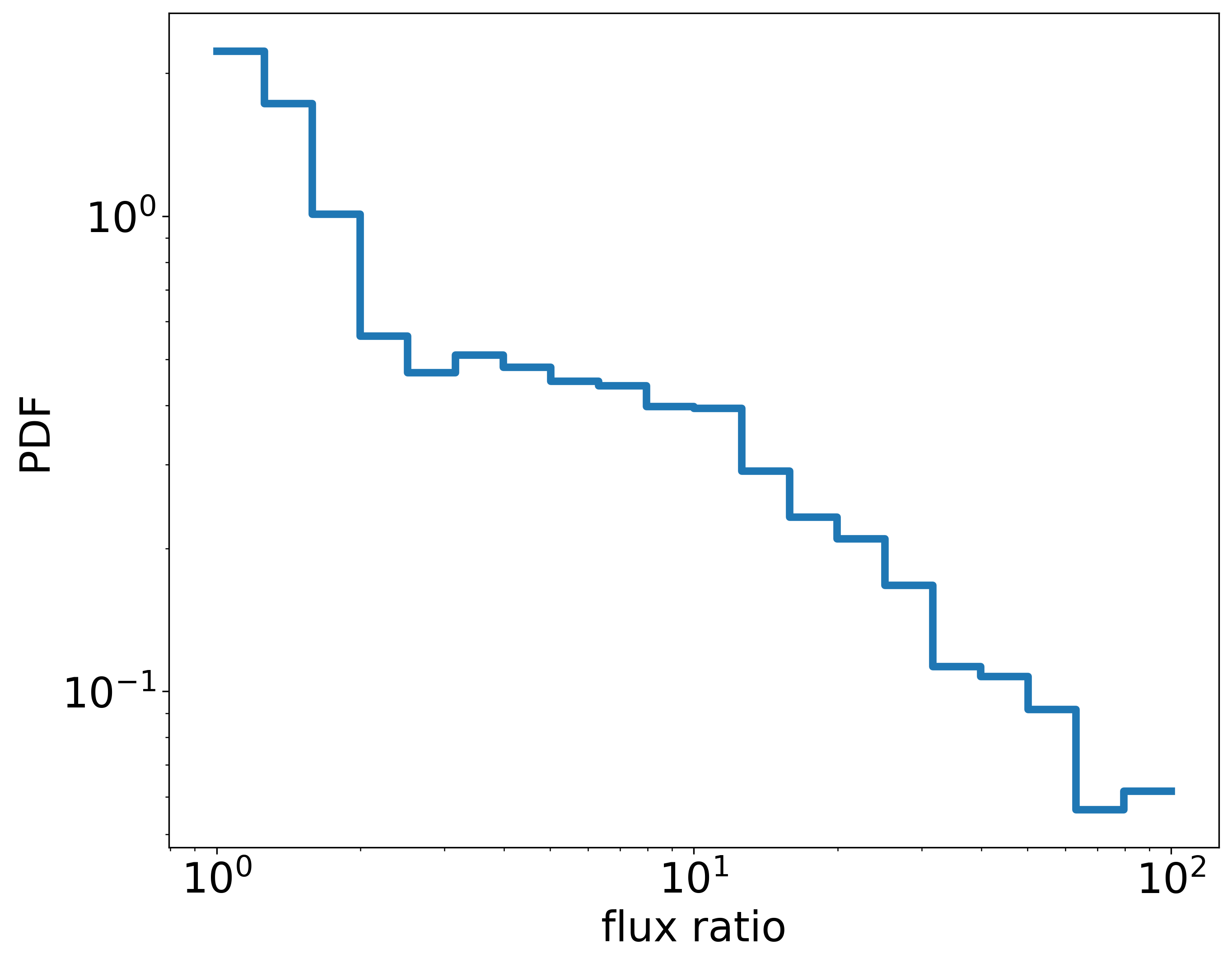}
    \caption{Properties of the mock strong gravitational lensing systems in the training samples. Upper left panel: The $i$-band apparent magnitude distributions for the lens galaxies, background quasars, and brightest lensed images, respectively. Upper right panel: The maximum angular separations between the multiple lensed images. Note that the separation range is truncated at 3$\arcsec$ in this plot. Lower left panel: The redshift distributions for the lens galaxies and background quasars. Lower right panel: The flux ratio between the brightest and second brightest lensed images.}
    \label{fig:properties_of_mock_lensed quasars}
\end{figure*}

We repeated these operations (randomly selecting the LRG and placing the quasar in the source plane) one million times. Consequently, the fluxes and positions of the multiple images were acquired and recorded into a catalog containing one million entries. We then applied selection criteria to isolate those with obvious strong lensing features. The criteria are as follows: first, the Einstein radius ($\theta_E$) must be greater than 0.6 arcsec, which is roughly the best FWHM of $r$-band seeing; second, the flux ratio between the brightest image and the foreground LRG must exceed 1. Here, we used the total fluxes of quasars and LRGs in the calculation. After applying these filters, a subset of 26,005 out of one million entries was selected. The distributions of the mock lensed quasars in the training sample are shown in Figure \ref{fig:properties_of_mock_lensed quasars}, including the lens and source redshifts, the flux ratio between the brightest and second brightest image, the separation, and the magnitudes of the sources and lens. Moreover, the median $\sigma_v$ is $222.80\ \mathrm{km/s}$, with a standard deviation of $42.92\ \mathrm{km/s}$; the quad to pair ratio is approximately $1:1.67$.

Based on these selected samples, we generated high-resolution ($0.05\arcsec$/pixel) ideal lensed images in $gri$-bands. We applied point spread functions (PSFs) to the ideal image plane, assuming the PSF follows the Moffat distribution:

\begin{equation}
\label{equ:moffat}
    I(r) = I_0 \left(1 + \left(\frac{r}{\alpha}\right)^2\right)^{-\beta}
    \label{equ:PSF-1}
\end{equation}
$I(r)$ is the intensity at a distance $r$ from the center of the distribution. $\alpha$ is the scale parameter related to the width of the distribution, and $\beta$ is the shape parameter, which determines the `heaviness' of the tails. Here, we fix $\beta$ at 2.2. Furthermore,
\begin{equation}
    \alpha = \frac{\text{FWHM}}{2 \cdot \sqrt{2^{\frac{1}{\beta}} - 1}}
    \label{equ:PSF-2}
\end{equation}
The detailed settings of the PSF are summarized in Table \ref{tab:PSF}. Consequently, we reduced the resolution of our mock data to align with the KiDS observational conditions ($0.2\arcsec$/pixel).

\begin{table}[htbp]
\centering
\caption{The PSF settings used to generate mock lensed quasars for the training and validation sets of the Network. Refer to Equations \ref{equ:PSF-1} and \ref{equ:PSF-2}, as well as the relevant discussion, for the interpretation of the listed parameters. The distribution aligns with the realistic distribution in KiDS DR4 \citep{Kuijken2019} and is consistent with those used in \cite{Li2021}.}
\label{tab:PSF}
\begin{tabular}{cccc}
\hline\hline
Parameter            & Range     & Units  & Distribution                                                \\ \hline
FWHM-g               & 0.6-1.2   & arcsec & normal ($\mu$=0.85, $\sigma$=0.1) \\
FWHM-r               & 0.50-0.90 & arcsec & normal ($\mu$=0.7, $\sigma$=0.05) \\
FWHM-i               & 0.55-1.2  & arcsec & normal ($\mu$=0.80, $\sigma$=0.1) \\
$\beta$ & 2.20      & -      & fixed                                                       \\
Axis ratio           & 0.98-1.0  & -      & uniform                                                     \\
Position angle       & 0-180     & degree & uniform                                                     \\ \hline\hline
\end{tabular}
\end{table}

We subsequently incorporated Poisson noise based on the low-resolution mocks of multiply-imaged quasars. The noise value $x$ obeys the Poisson distribution:
\begin{equation}\label{equ:poi_nosie}
    f(x)=\frac{\lambda^x}{x!}e^{-\lambda}
\end{equation}
where $\lambda$ is the ADUs at a certain pixel position of the noiseless image. The ADUs can be calculated by Gain * Flux * Exp\_time, where Gain, a factor that converts the number of electrons into the number of ADUs, can be obtained from the FITS header file of KiDS observation images. The exposure time varies across different bands\footnote{https://kids.strw.leidenuniv.nl/DR4/index.php} (1800 seconds for the $r$ band, 900 seconds for the $g$ band, and 1200 seconds for the $i$ band). Note that the Poisson noise was not added to the lensing galaxy, since their images were taken from observations and the noise already included.

For the negative part of the training set, we randomly selected objects in the predictive dataset, ensuring that no known lenses were included in this process. To balance the negative samples with the positive ones, we made the volume of each set approximately the same. Thus, 27,275 negative examples were selected.

\begin{figure*}[htbp]
    \centering
    \includegraphics[scale=0.55]{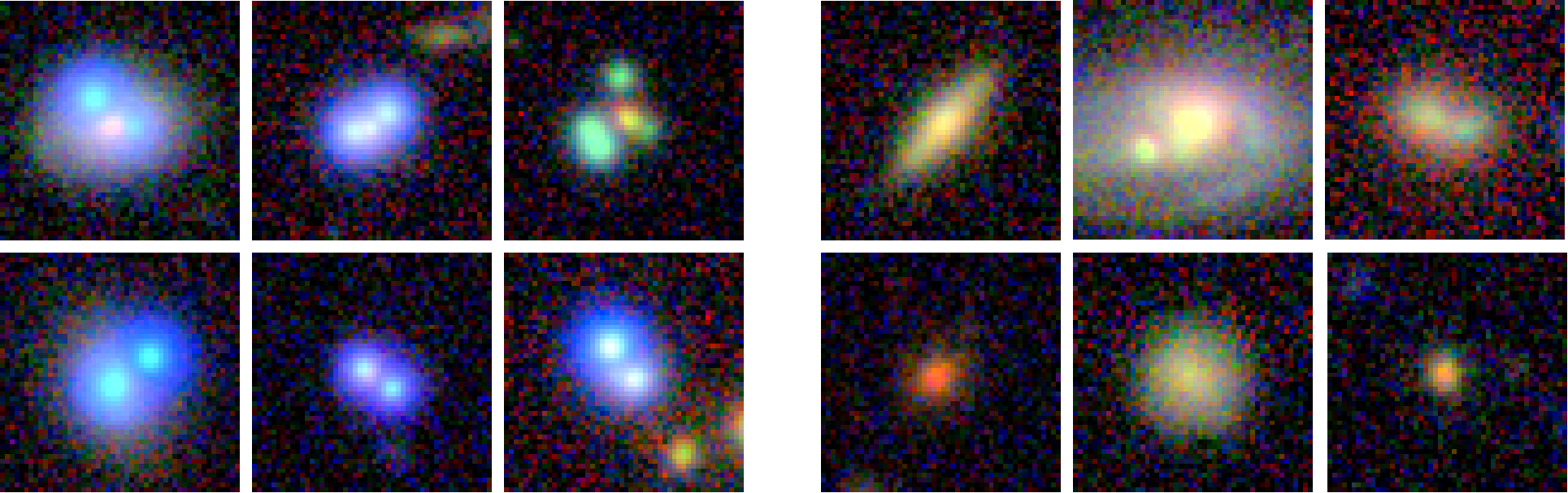}
    \caption{Examples of the training samples used in this work, presented as $gri$ 3-band composite images. The left panel shows the mock lensed quasars (positives), while the right panel displays the non-lensed objects (negatives). Each cutout image has a size of 10$\arcsec$×10$\arcsec$.}
    \label{fig:train_sample_demo}
\end{figure*}

In summary, we constructed the training and validation sets by focusing on the portion of the mock catalog that exhibits the most distinct strong lensing features of lensed quasars. In Figure \ref{fig:train_sample_demo}, we show some examples of the positive and negative samples in the training set. In total, we created 26,005 mock lensed quasars as the positive set, complemented by 27,275 real non-lenses as the negative set. We allocated 60\% of these datasets for training the Network, 10\% for validation purposes, and reserved the remaining 30\% for testing the selection effect of the Network (named as simulated testing set).

\subsection{Testing Set}
\label{sec: test set}

To evaluate the efficacy of our classifier, we created two testing datasets incorporating both positive and negative instances. The first dataset, referred to as the simulated testing data, was introduced in Section \ref{sec:training set}; the second dataset, named the real test set, included negatives directly sourced from the target dataset and positives from known lensed quasars with applied data augmentation.

\begin{figure*}[htbp]
    \centering
    \includegraphics[scale=0.65]{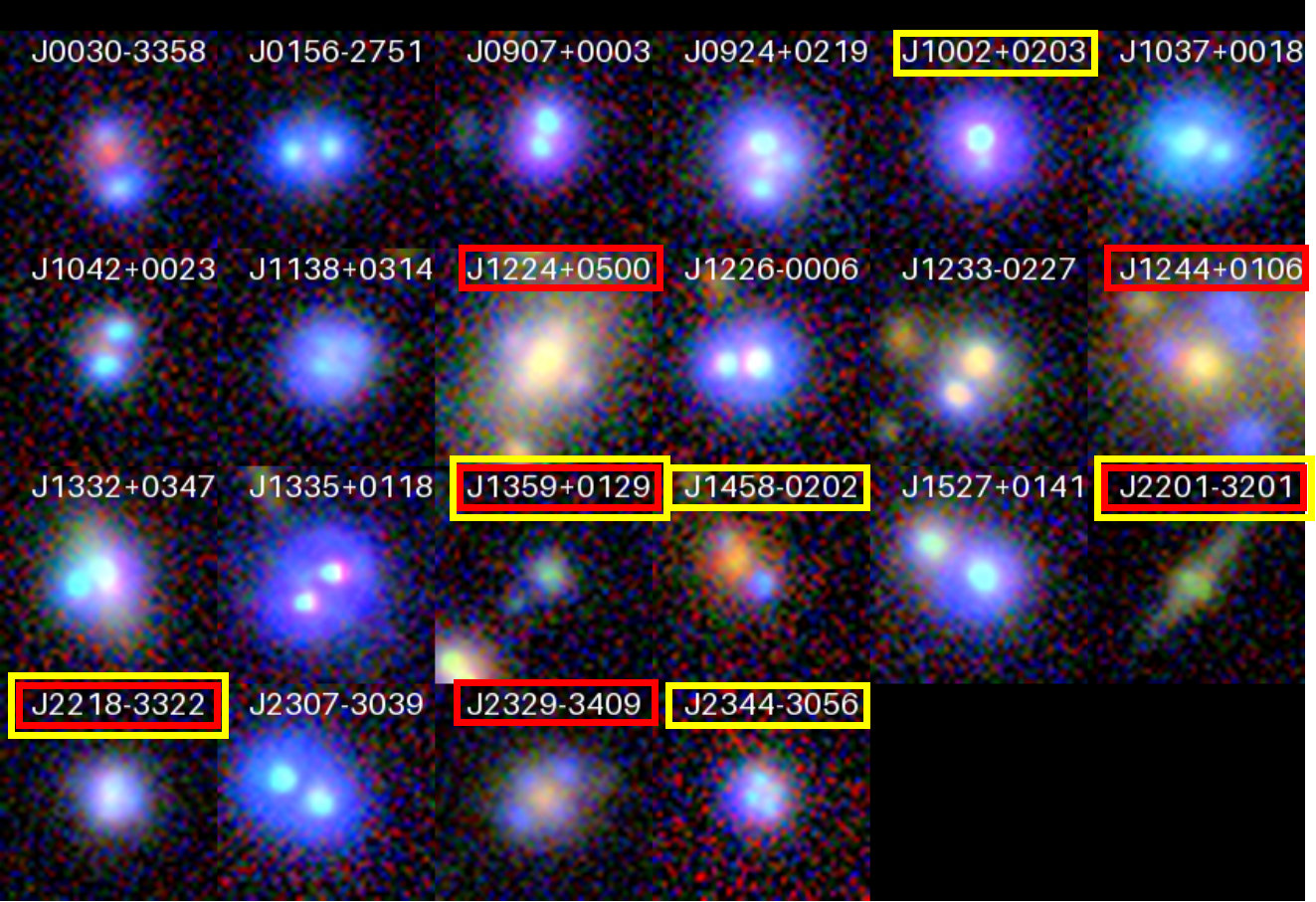}
    \caption{The confirmed lensed quasars and lensed compact galaxies in KiDS footprint. The ones marked by red frames were not included in the test set. The ones marked by yellow frames were missed by the Network. The cutout size is 8.8$\arcsec$ × 8.8$\arcsec$. The detailed information is given in Table.\,\ref{tab:real_lensed quasars}. Left is east, up is north.}
    \label{fig:real_lensed quasars}
\end{figure*}

For the real test set, the negative subset comprised 50,000 samples randomly drawn from the target data, with rigorous visual scrutiny confirming the absence of any strong lensing features. The positive subset was identified by cross-matching the target data with the SLED database\footnote{C. Lemon, private communication; SLED, Vernardos et al., in preparation}, resulting in 22 known lensed quasars and lensed compact galaxies. Figure \ref{fig:real_lensed quasars} displays all the cutouts from our cross-matching, while Table \ref{tab:real_lensed quasars} lists the names of these lensing systems alongside their respective references. Notably, we used the lensed quasars as our positive subset, excluding lensed galaxies due to their more extended nature (J1244+0106, J2329-3409) or significantly different color (J1224+0050) compared to lensed quasars. Additionally, three systems (J1359+0129, J2201+3201, J2218-3322) with atypical image features in KiDS imaging were eliminated. Thus, six out of the 22 strong lenses were excluded, leaving 16 systems for the test set. These 16 systems underwent the following data augmentation process to better evaluate performance.

Each image in the real test set was subjected to a series of transformations, including vertical flipping, horizontal flipping, diagonal flipping, and random rotation within the range of 0-360 degrees (applied twice). Consequently, each image was enhanced through five repetitions of these operations, with the fourth operation applied twice. This process increased the dataset's volume sixfold, encompassing both the original images and their augmented counterparts. As a result, we constructed a test set comprising 50,000 negatives and 96 positives (16 original positives augmented sixfold). Notably, in this work, the data augmentation process was only used in the real test set and not applied during the training process.

\begin{table*}
\tiny
\centering
\caption{The detailed information of the cutouts in Figure \ref{fig:real_lensed quasars} is provided above. J1224+0500, J1244+0106, and J2329-3409 are lensed galaxies. `Rediscovered' indicates whether this system had earned a $p_{CNN}$ larger than the threshold of 0.94. `In TestsingSet' indicates whether this system was used in the test set, denoted by `Y' for yes and `N' for no. `D23' refers to \cite{Dawes2023}, `L23' refers to \cite{Lemon2023}, and `H23' refers to \cite{He2023}.}
\label{tab:real_lensed quasars}
\begin{tabular}{cccccccccc} 

\hline\hline
name                    & RA (J2000) & Dec (J2000) & image separation (arcsec) & $z_s$ & $z_d$  & $n_{\rm img}$ & Rediscovered & In TestsingSet & References   \\ 
\hline
J0030-3358              & 7.674      & -33.9767    & 2.03                      & 1.58  & 0.715  & 2      & Y            & Y              & \cite{Huang2021},L23         \\
J0156-2751              & 29.1039    & -27.8562    & 1.5                       & 2.97  &        & 2      & Y            & Y              & D23,H23,L23  \\
J0907+0003              & 136.7937   & 0.0559      & 1.32                      & 1.299 &        & 2      & Y            & Y              & \cite{Spiniello2019}             \\
SDSSJ0924+0219          & 141.23246  & 2.32358     & 1.34                      & 1.685 & 0.393  & 4      & Y            & Y              & \cite{Inada2003}             \\
CXCOJ100201.50+020330.0 & 150.5063   & 2.0581      & 0.85                      & 2.016 & 0.439  & 2      & N            & Y              & \cite{Jaelani2021}             \\
J1037+0018              & 159.3665   & 0.3057      & 1.25                      & 2.462 &        & 2      & Y            & Y              & \cite{Lemon2022}             \\
KIDS1042+0023           & 160.6553   & 0.3839      & 1.45                      & 2.26  &        & 2      & Y            & Y              & \cite{Spiniello2019}            \\
SDSSJ1138+0314          & 174.51554  & 3.24939     & 1.44                      & 2.438 & 0.445  & 4      & Y            & Y              & \cite{Inada2008}             \\
J1224+0500              & 186.233403 & 0.846681    &                           & 1.10  & 0.2372 & 4      & Y            & N              & \cite{Napolitano2020}             \\
SDSSJ1226-0006          & 186.5334   & -0.10061    & 1.21                      & 1.125 &        & 2      & Y            & Y              & \cite{Inada2008},H23         \\
J1233-0227              & 188.4219   & -2.4604     & 1.76                      & 1.598 & 0.345  & 2      & Y            & Y              & D23,L23      \\
J1244+0106              & 191.21402  & 1.111979    &                           & 1.069 & 0.388  & 4      & Y            & N              & \cite{Stark2013}             \\
SDSSJ1332+0347          & 203.09425  & 3.79442     & 0.99                      & 1.685 & 0.191  & 2      & Y            & Y              & \cite{Morokuma2007}             \\
J1335+0118              & 203.894979 & 1.301553    & 1.62                      & 1.57  & 0.44   & 2      & Y            & Y              &  \cite{Inada2008}            \\
J1359+0129              & 209.9332   & 1.47053     & 1.05                      & 1.096 &        & 2      & N            & N              & \cite{Khramtsov2019}             \\
SDSSJ1458-0202          & 224.6982   & -2.03497    & 2.15                      & 1.724 &        & 2      & N            & Y              & \cite{More2016}             \\
ULASJ1527+0141          & 231.8338   & 1.69433     & 2.55                      & 1.439 & 0.3    & 2      & Y            & Y              & \cite{Jackson2012}             \\
CY2201-3201             & 330.38667  & -32.02889   & 0.83                      & 3.9   & 0.32   & 2      & N            & N              & \cite{Castander2006}             \\
J221849.86-332243.6     & 334.7075   & -33.3787    & 0.6                       & 2     &        & 4      & N            & N              & \cite{Chen2022}             \\
KIDS2307-3039           & 346.8287   & -30.6547    &                           & 2.64  & 0.46   & 2      & Y            & Y              & \cite{Spiniello2019}             \\
J2329-3409              & 352.417763 & -34.156397  &                           & 1.59  & 0.3810 & 4      & Y            & N              &  \cite{Napolitano2020}            \\
WISE 2344-3056          & 356.0708   & -30.9405    &                           & 2.18  & 1.298  & 4      & N            & Y              &  \cite{Schechter2017}            \\
\hline\hline
\end{tabular}
\end{table*}

\section{Methodology} \label{sec:method}

In Section \ref{sec: cnn and metric}, we present the Network architecture and elaborate on its evaluation metrics. Using the aforementioned training and test sets (Sections \ref{sec:training set} and \ref{sec: test set}), the Network was trained, and the Network parameters are fine-tuned to achieve the best performance in the real test set. This is detailed in Section \ref{sec:train test}.

\subsection{Network and Evaluation Metric}
\label{sec: cnn and metric}

We used a ResNet-18 \citep{He2016} as our classifier. This Network has been adopted by plenty of lens-finding works \citep[e.g.][]{Lanusse2018,Huang2021,Li2020,Li2021}. The inputted data is three-band $gri$ fits with a square root applied to it, the normalization is applied in each band separately. The shape of the inputted data is 50 pixel × 50 pixel ×3 channel. The Network was trained from scratch, with no pre-training process applied. The detailed Network architecture can be found at this link\footnote{\url{https://github.com/EigenHermit/H24/blob/main/Network-structure}}.

Classic metrics were used for evaluating the performance of Networks. Firstly, the TPR (true positive rate, or recall) is defined by:
\begin{equation}
{\rm TPR} = \frac{{\rm TP}}{{\rm TP}+{\rm FN}},
\end{equation}
where TP represents the lensed quasars correctly picked out by the classifier. FN represents the lensed quasars mistakenly classified by the Network. Thus, TP+FN is the overall number of lensed quasars that exist in the testing dataset. TPR measures how well the classifier can recover the lensed quasars. Secondly, the false positive rate (FPR) is defined by:
\begin{equation}
{\rm FPR} = \frac{{\rm FP}}{{\rm TN}+{\rm FP}},
\end{equation}
where FP means the non-lensed quasars incorrectly labeled as lensed quasars by the Network, and TN represents the non-lensed quasars correctly labeled. Last but not least, the $accuracy$ measures the overall performance of our classifier:
\begin{equation}
{\rm accuracy} = \frac{{\rm TP+TN}}{{\rm TP+FP+TN+FN}},
\end{equation}
Furthermore, the ROC (receiver operating characteristic) curve was also adopted in this work to measure the capacity of the Network when choosing different thresholds ($p_{th}$) of the Networks. The $x,y$-axes of the ROC curve are TPR and FPR. The data points on the curve represent the Network's TPR and FPR when choosing different thresholds of the Networks.

\subsection{Training and Testing of the Network}
\label{sec:train test}

With the training and validation sets obtained in Section \ref{sec:training set} and the network described in Section \ref{sec: cnn and metric}, we proceeded to train the Network. The training loss was computed on the training set, while the validation loss was computed on the validation set. Subsequently, we fine-tuned the number of epochs ($ep$) and learning rate ($lr$) based on two key criteria.

\begin{figure*}[htbp]
    \centering
    \includegraphics[scale=0.5]{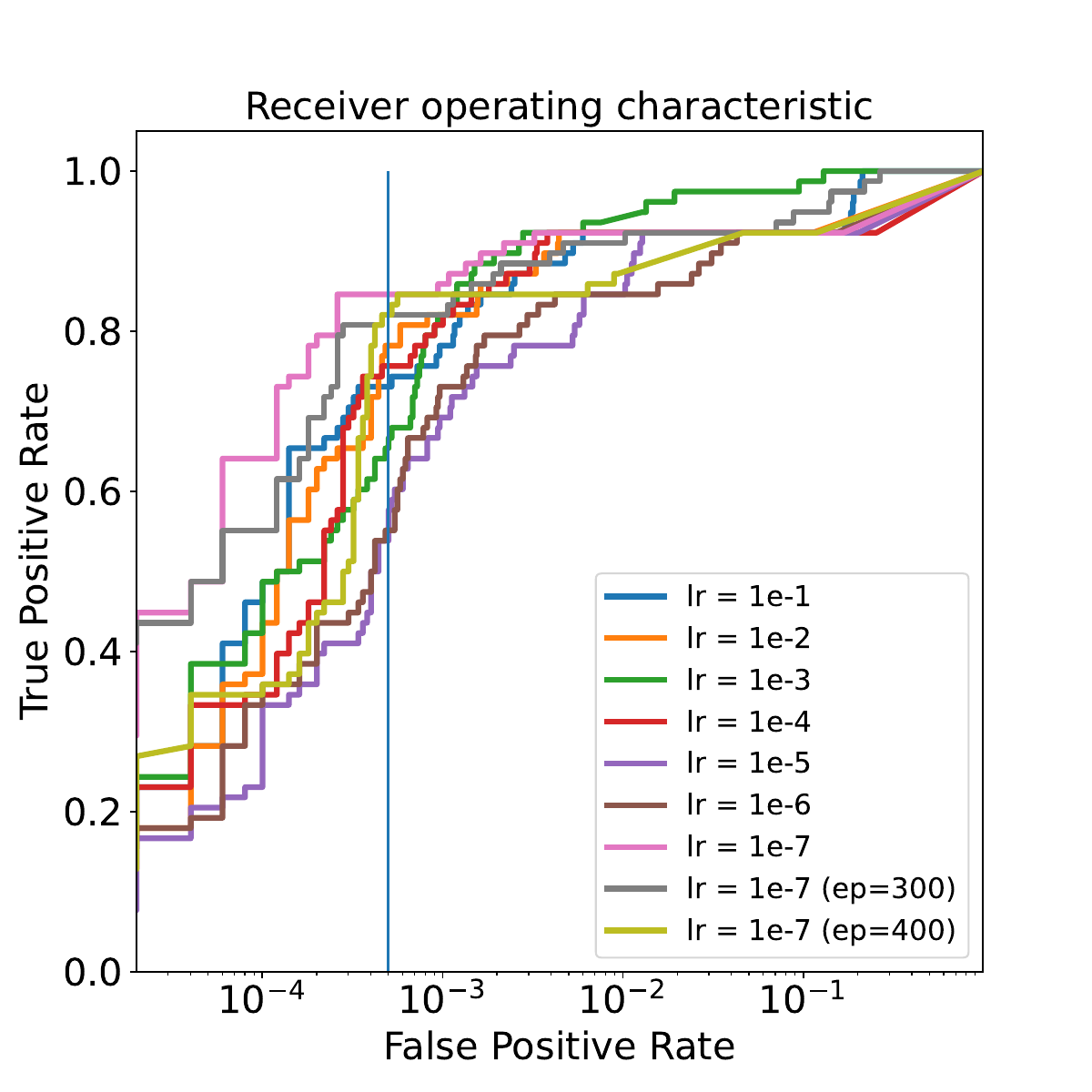}
    \caption{The figure shows the fine-tuning process of Network. `lr' represents the learning rate, `ep' stands for the total number of epochs that were used in the training processes. The blue solid vertical line shows the FPR equals 1/2000.}
    \label{fig:fine-tuning}
\end{figure*}

Firstly, it is crucial for both the training loss and validation loss to converge at a similar pace. Secondly, we assessed the Network's performance on the real test set, focusing particularly on the TPR at an FPR of approximately 1/2000. We chose this FPR because, given the volume size of the data to be applied (approximately 10 million), an FPR of around 1/2000 corresponds to a candidate size of roughly 5000, which we found suitable for subsequent human inspection. The Network's performance under the tested parameters is depicted in Figure \ref{fig:fine-tuning}. If not labeled in Figure, the $ep$ was set to 200. Notably, when trained with longer epochs, the TPR at 1/2000 shows a decrease; see the $ep=300$ and $ep=400$ lines for this result, which may contributed by the over-fitting.

\begin{figure}[htbp]
    \centering
    \includegraphics[scale=0.55]{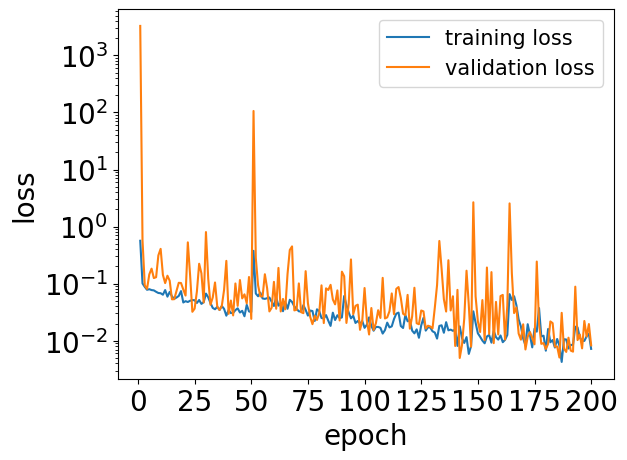}
    \caption{Loss value evaluated on the training and validation samples respectively. We set the save\_best = True, which means the model with lowest validation loss will be saved, rather than the model from the last epoch. Thus the epoch = 186 model was used. See more discussion in Section \ref{sec:train test}.}
    \label{fig:loss}
\end{figure}

Upon evaluation, setting save\_best = True ensures that the training program continuously saves the model with the lowest validation-loss as the final model, rather than relying on the model from the last epoch. The optimal parameters identified were a learning rate of $1 \times 10^{-7}$ and a total of 186 epochs. The loss value for this network is illustrated in Figure \ref{fig:loss}. With these settings, the achieved true positive rate (TPR) is 0.8125, using a threshold of $p_{th} = 0.94$. This threshold will be used in subsequent stages of the application process. It is important to note that the limited number of known lensed quasars in the real test set introduces Poisson noise into the evaluation, despite the application of data augmentation. Nevertheless, performance with real data is preferable to that obtained from a simulated test set, as the Network tends to perform exceptionally well on simulated data. Additionally, results based on real data are more meaningful due to potential discrepancies between actual lensed quasars and their simulated counterparts.

During the fine-tuning process, we serendipitously encountered a nice quad candidate, which is detailed in Appendix \ref{sec:app}.

\section{Results}

We applied the best-performance Network (Section \ref{sec:train test}) to the target dataset (Section \ref{sec:tobeapplied_data}). The Network identified 4,195 candidates assuming a $p_{th}=0.94$. Following this, the first author Z.H. inspected them, discerning and eliminating 1,057 highly improbable objects, including clearly bright stars, extended objects, leaving us with the remaining candidates undergo thorough human scrutiny involving ten individuals (Section\,\ref{sec:human_ins}). Finally, the final candidates are selected and categorized into high-, median-, lower-score candidates (Section \ref{sec:cand_list}) based on their average scores received in the human inspection process. We also compared our results with other candidate lists in SLED.

\label{sec:result}

\subsection{Human Inspection}
\label{sec:human_ins}

Ten authors were involved in the scoring process, which contains three rounds. The adopted scheme is as follows:

\begin{itemize}
\item Score 3 - The object has clear multiple point-like images and a central galaxy, and a lens model can easily reproduce their configuration.
\item Score 2 - Case 1: Clear multiple point-like (with similar color) images but without a clear central galaxy. Case 2: There are multiple point-like images and a central galaxy, but the configuration is atypical.
\item Score 1 - Case 1: Likely to be multiple point-like images without a clear central galaxy, while the color is less similar (compared to the color similarity of Case 1 in ‘Score 2’). Case 2: There is a clear central elliptical galaxy and one point-like source located near the galaxy but lacking other counter images.
\item Score 0 - Case 1: The color of two point-like sources differs largely or they are located distantly. Case 2: Everything else (like bright stars, disk galaxies, mergers, artifacts, etc.).
\end{itemize}

In the first round, or the calibration round, 200 common objects, including 12 known lensed quasars and 4 lensed compact galaxies that were successfully rediscovered by the Network (see Figure \ref{fig:real_lensed quasars}), were graded by ten authors. We used $S_{\text{av}}$ to denote the average scores an object received, and $var(S_{\text{av}})$ to denote the variance of $S_{\text{av}}$. When the systems received different opinions (the $var(S_{\text{av}}) > 0.8$) and the confirmed systems received at least one zero score, they were discussed to cross-calibrate our grading criteria and then achieve a consensus of the grading criteria for the 2nd and 3rd round grading. Final adopted criteria have been listed above. Here, we choose 0.8 to select the most controversial candidates and control the sample size that we need to discuss.

For the second round, the 3,172 candidates were evenly distributed among the authors. Each member scored 386 candidates, which contains $\sim 69$ systems from 200 objects that were already inspected in the first round to check the consistency. 

\begin{figure}
    \centering
    \includegraphics[scale=0.4]{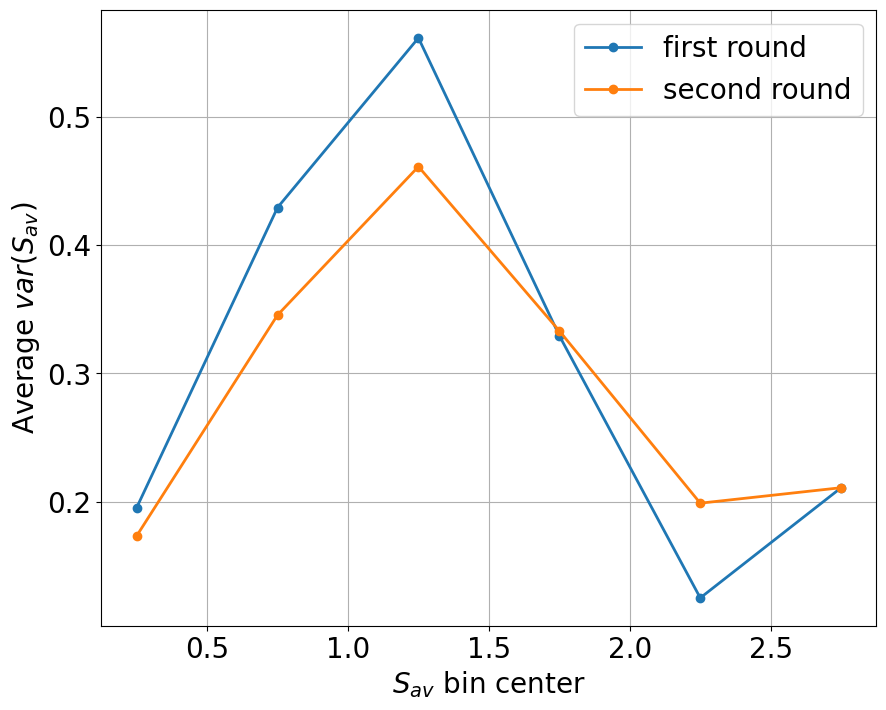}
    \caption{The $S_{\text{av}}$ versus average variance of $S_{\text{av}}$ in respective bin of 200 common objects, the bin size is 0.5. Those objects were used to check the consistency of inspectors in the first and second round of human grading.}
    \label{fig:sav}
\end{figure}

In Figure \ref{fig:sav}, we compare the distribution of $S_{\text{av}}$ for the 200 common objects with the first-round results. For objects that are clearly non-lenses, with low $S_{\text{av}}$, the inspectors agree, and the scatter in the scores is small (around 0.2). Similarly, for the most obvious lenses, with high $S_{\text{av}}$, all inspectors tend to agree, and the scatter is again small, around 0.1-0.2. A different story emerges for the remaining inspected objects, those around a score of 1, which are more uncertain and show greater variability in the inspectors' decisions.

Overall, we observe that the $var(S_{\text{av}})$ is reduced in this round. We then recorded the scores for each candidate, and for the 200 systems that received multiple scores, we calculate the average scores. Based on these (average) scores, we eliminate the objects that received a zero score. This resulted in the rejection of 1,297 candidates. 

In the third round, the remaining 1,879 objects were evenly distributed the ten inspectors, with each member responsible for inspecting 374 or 376 objects. Combining the grades from the second and third rounds, each of the 1,879 objects received at least three ratings.

After the inspection process, the $S_{\text{av}}$ for each object was determined. The interval selections were established with reference to \cite{Shu2022}. Objects scoring between [2.33, 3] were designated as high-score candidates. Those scoring between [1.67, 2.33) were categorized as median-score candidates, while objects with $S_{\text{av}}$ between [1.33, 1.67) were classified as lower-score candidates. Any $S_{\text{av}}$ below 1.33 resulted in the candidate being disqualified.
Among the 12 known lensed quasars and 4 lensed compact galaxies that underwent human inspection, eight were classified as high-score systems, while the remaining eight were classified as median-score systems. The average score of the 16 systems is 2.44.

It is worth noting that we decided not to use the Grade A, B, C classifications commonly employed in galaxy-galaxy lensing studies. This is because when identifying strongly lensed quasars candidates, it is challenging to eliminate star contaminants in the candidates, especially in doubles. In contrast, in the case of lensed galaxies, the distorted galaxies are easier to distinguish from other possible contaminants, such as ring galaxies and mergers.

\subsection{Final candidates}
\label{sec:cand_list}
\begin{figure*}[htbp]
    \centering
    \includegraphics[scale=0.75]{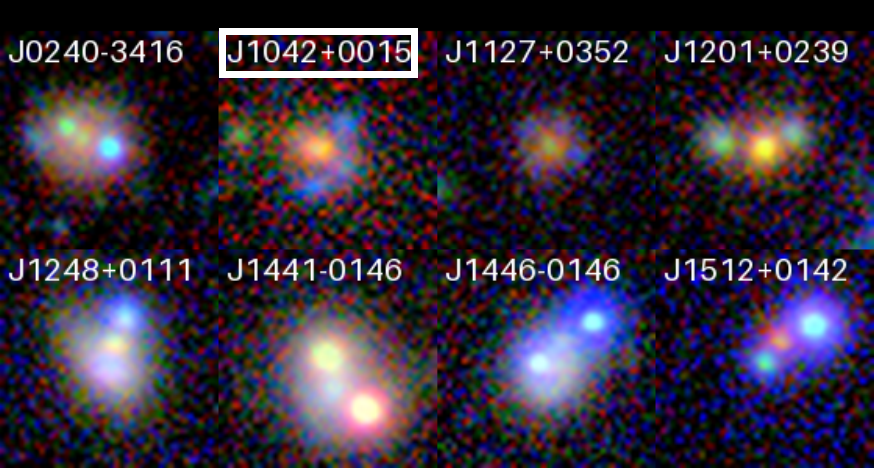}
    \caption{Cutouts of the identified high-score systems (excluding confirmed lenses, which are displayed in Figure \ref{fig:real_lensed quasars}) are presented here. The left side represents east, while the top indicates north. A white frame signifies that the system has been identified in other studies, as detailed in Table \ref{tab:H24}, under the column labeled SLED\_xmatch. Each cutout measures 8.8 × 8.8 arcsec$^2$.}
    \label{fig:Acand}
\end{figure*}

\begin{table*}[]
\caption{Descriptions of the properties provided for each lensed quasar candidates in our final catalog. The catalog contains new candidates and includes rediscovered confirmed systems. The full catalog is available online (see the footnote in the abstract).}
\label{tab:H24}
\centering
\begin{tabular}{cc}
\hline\hline
\multicolumn{2}{c}{\textbf{All candidates}}                                                   \\ \hline
\multicolumn{2}{c}{\textit{This catalog contains candidates identified in this work}}     \\ \hline
Columns              & Description                                                            \\ \hline
name                 & Name of this system, generated automatically by its RA, dec            \\
RA                   & Right Ascension (J2000)                                                \\
Dec                  & Declination (J2000)                                                    \\
grade                & Grades of lensed quasar candidates (high, median, lower-scores)                           \\
$n_{\rm img}$        & Number of images                                                       \\
separation           & The image (maximum) separation of this system                              \\
SLED\_xmatch         & Cross-match information with SLED  database.                            \\
                     & There are three kinds of Label: confirmed, contaminant, or candidates. \\
\multicolumn{1}{l}{} & Labels are followed by relevant reference(s).                          \\                                        
$S_{\text{av}}$      & The average score received in human inspection                         \\
$p_{\text{Network}}$ & The probability given by the Network                                   \\
isKnownLenses        & Whether this object exist in Table \ref{tab:real_lensed quasars}                  \\ \hline\hline
\end{tabular}

\end{table*}

\begin{table*}[]
\caption{Descriptions of the properties given for each images of each candidate lensed quasar in the detailed catalog. The full catalog is available online (see the footnote in abstract).}
\label{tab:H24-details}
\centering
\begin{tabular}{cc}
\hline\hline
\multicolumn{2}{c}{\textbf{Candidates in detail}}                                                 \\ \hline
\multicolumn{2}{c}{\textit{This catalog describes every images of each candidate}} \\ \hline
Columns        & Description                                                                      \\
name           & Name of this candidate system that this image belongs to                         \\
RA             & Right Ascension (J2000) of the center of this image                              \\
Dec            & Declination (J2000) of the center of this image                                  \\
specz                & Spectroscopy redshift from public available data-sets                  \\
specz\_cite          & Sources of `specz' of this object from SLED database                   \\
PM\_SIG        & Proper motion significance defined in \cite{Lemon2018}                           \\
PLX\_SIG       & Parallax motion significance defined in \cite{Lemon2018}                         \\
$u$            & $u$-band magnitude of this image                                                 \\
$g$            & $g$-band magnitude of this image                                                 \\
$r$            & $r$-band magnitude of this image                                                 \\
$i$            & $i$-band magnitude of this image                                                 \\
photo-$z$      & The quasar machine learning redshift obtained using GaZNet \citep{Li2022GaZNet}    \\
grade          & The grades (high, median, lower-scores) of the candidate system that this image belongs to.           \\ \hline\hline
\end{tabular}
\end{table*}

Following the human inspection process outlined in Section \ref{sec:human_ins}, we find 272 candidates that have received score higher than 1.33, thus they were selected by our study. 

To illustrate them, we build two catalogs. The first one provides a comprehensive overview of the catalog's contents, including the following information: RA, Dec, grade by human inspection, number of images, maximum separation, SLED cross-matching results, the average scores from human inspection, and a flag indicating whether the system is in the test set. The detailed explains about those columns can be found at Table \ref{tab:H24}. On the other hand, Table \ref{tab:H24-details} offers a more detailed breakdown of each image. This detailed table includes the RA and Dec of each image, spectroscopic redshifts and their sources, astrometric information such as $\tt{PM\_SIG}$ and $\tt{PLX\_SIG}$, the $u$, $g$, $r$, and $i$ magnitudes provided by KiDS, and the photo-$z$ of each image as determined by GaZNet \citep{Li2022GaZNet}.

\begin{figure*}[htbp]
    \centering
    \includegraphics[scale=1]{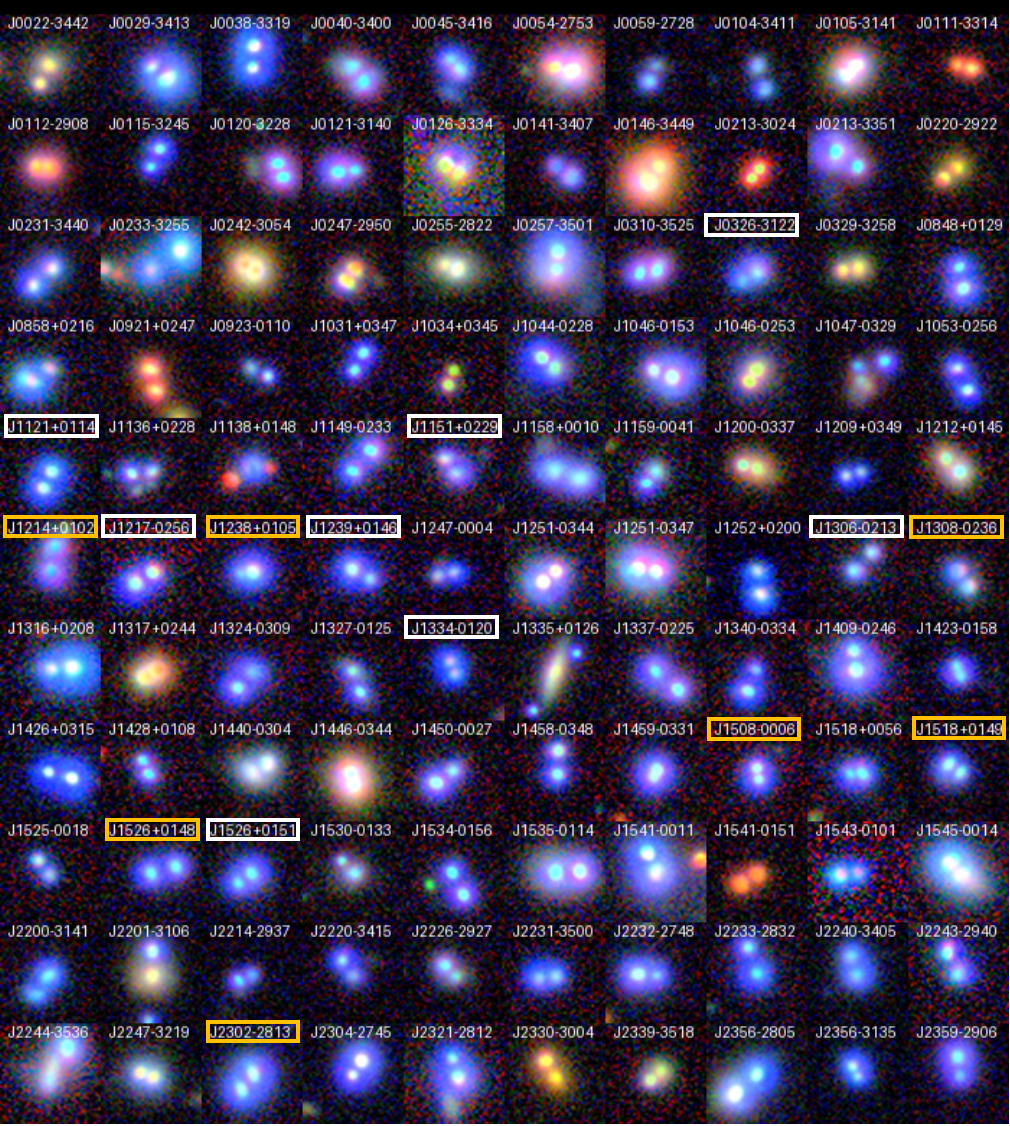}
    \caption{Cutouts of identified median-score systems are displayed. Each cutout is sized at 8.8$\arcsec$ × 8.8$\arcsec$. Left is east, up is north. A orange frame denotes reported contaminants in other studies. Candidates that have been identified in other studies (but have not been spectroscopically observed) are marked by white frames. Refer to column `SLED\_xmatch' in Table \ref{tab:H24} for details about contaminants or overlaps with other works.}
    \label{fig:Bcand}
\end{figure*}

According to the classification standards defined in Section \ref{sec:human_ins}, our catalog includes 16 high-score candidates, 118 median-score candidates, and 138 lower-score candidates. After excluding systems confirmed through spectroscopy, the revised counts are 8 high-score, 110 median-score, and 138 lower-score candidates. The color images of these candidates are presented in Figures \ref{fig:Acand}, \ref{fig:Bcand}, and \ref{fig:Ccand}. Median-score and lower-score candidates will undergo PSF subtraction to determine if any lensing galaxies are visible, which is crucial for further assessing their potential as lenses. The results will be presented in a separate publication.

Cross-matching with SLED revealed that some candidates and contaminants had been previously reported. Candidates that were reported earlier without further spectroscopic confirmation are highlighted with white frames, while systems that have undergone spectroscopic observation and were confirmed as non-lenses are marked with orange frames.
After excluding these candidates and contaminants, the final counts are as follows: 7 high-score candidates, 95 median-score candidates, and 127 lower-score candidates, resulting in a total of 229 candidates. Specifically, the high-score candidates have only one overlap, which has not yet been subjected to spectroscopic observation. The median-score candidates exhibit 15 overlaps with SLED, of which 7 are contaminants, yielding a contaminant rate of approximately 47\%. For the lower-score candidates, there are 11 overlaps, with 8 identified as contaminants, resulting in a contaminant rate of roughly 73\%.
This suggests that the purity of the median-score candidate list is higher than that of the lower-score candidates, although the statistics are affected by significant Poisson noise due to the small number of overlaps. More precise purity assessments can be made once spectroscopic observations are conducted.

We found that 40 out of 272 systems have at least one spectroscopic redshift detection by other spectroscopic surveys like Sloan Digital Sky Survey \citep[SDSS,][]{Blanton2017} and Dark Energy Spectroscopic Instrument Early Data Release \citep[DESI-EDR,][]{DESIEDR} and targeted observations like \cite{Lemon2022}. In the final catalog, these redshifts are recorded in the `specz' column, with the sources of the spectroscopic redshifts indicated in the `cite\_z' column (Table \ref{tab:H24-details}).

In Figure \ref{fig:cand_pro}, we show the distributions of various parameters: quasar spectroscopy redshift of 40 systems that have spectroscopic detections, $i$-band magnitudes, image (maximum) separation ($sep$), and $g-r$ vs. $r-i$. Note that the $i$-band magnitudes and $g-r$ vs. $r-i$ plots pertain to the quasar images, excluding the lensing galaxies. 

The 272 candidate lensed quasars in the study span a redshift range of [0.422, 3.57], with a peak around 1.5. Their $i$-band apparent magnitudes range from $16 \leq i \leq 22$, with a peak around $i \sim 20$. This distribution reveals that the luminosity distribution of the quasars in this work deviates from the power-law trend given by \cite{Oguri2010} around $i$=20, indicating that the method tends to select samples at lower redshifts ($<$2.0) and misses high-redshift lensed quasars. This is likely due to the training sample being dominated by lower redshift quasars, as shown in Figure \ref{fig:properties_of_mock_lensed quasars}.
Furthermore, the spectroscopic redshifts are primarily from SDSS, which is limited by its own selection effects.

The image (maximum) separation ranges from $0.8 \leq sep \leq 2.75$ arcseconds, peaking at approximately 1.5 arcseconds. This is largely due to the training sample setting, in which $sep$ peaks at 1.3\arcsec. The $g-r$ vs. $r-i$ color-color plot shown in Figure \ref{fig:cand_pro} reveals that high-score candidates are more consistent with known lensed quasars, while the ones circled by the pink circle has a higher risk of being star contamination.

According to \cite{Oguri2010}, by selecting an $i$-band limit magnitude of 23.6 and considering the 1,347 square degree survey area of KiDS DR5, approximately 170 lensed quasars are expected to be found within KiDS DR5, with a predicted quad-to-pair ratio of about 1/6. Previous studies have identified 19 lensed quasars, suggesting that around 151 systems are still yet to be discovered. In our study, we identified 229 new candidates that can contribute to the pool. As shown in Figure \ref{fig:cand_pro}, our sample is expected to contribute significantly to the $i$-band magnitude range of 20 to 21. 


Taking both the rediscovered lenses and candidates into consideration, our findings reveal a quad-to-pair ratio of 1/66, significantly lower than the ratio in the training set, which is approximately 1/1.67, and also lower than the predicted ratio of 1/6 by \cite{Oguri2010}. This discrepancy can be attributed to two factors.
Firstly, based on known lensed quasars and the Network’s predictions, it appears that about half of the quads are missed, while approximately one-quarter of known pairs are overlooked. This suggests that quads are more likely to be missed by the Network.
Secondly, the false positive rate for pair candidates is expected to be higher than that for quads, as pairs are more prone to being mimicked by stars, dual quasars, and projected quasars.

\begin{figure*}[htbp]
    \centering
    \includegraphics[scale=1.1]{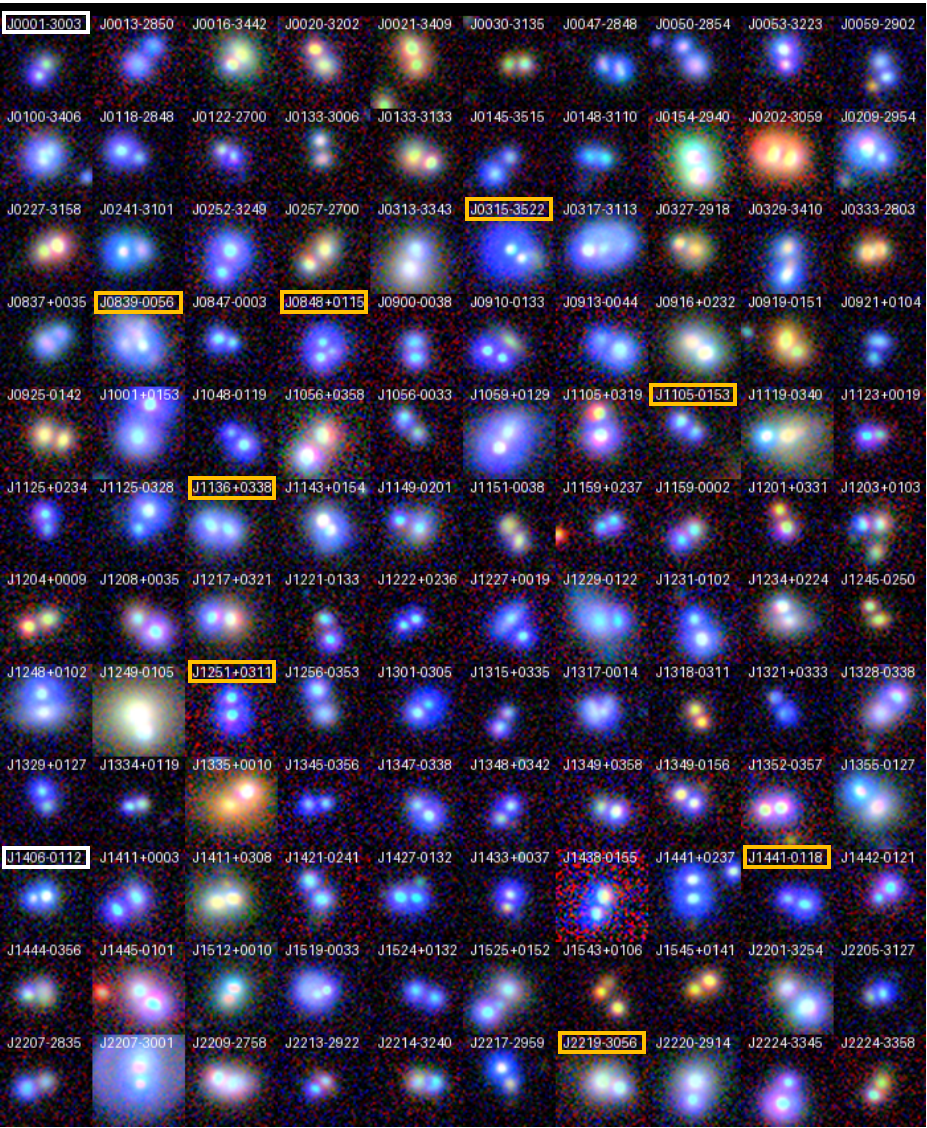}
    \caption{Cutouts of identified lower-score systems are presented. The color scheme of the labels, the direction, and cutout size remain consistent with those in Figure \ref{fig:Bcand}.}
    \label{fig:Ccand}
\end{figure*}

\addtocounter{figure}{-1}
\begin{figure*}[htbp]
    \centering
    \includegraphics[scale=1.1]{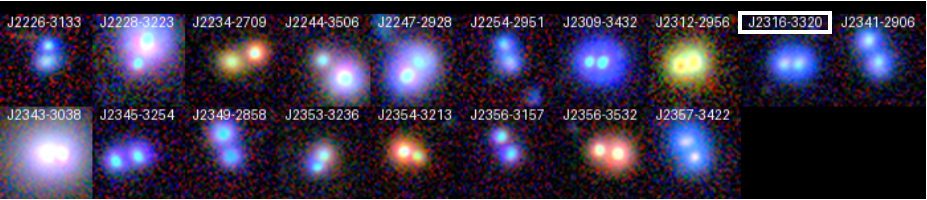}
    \caption{Continued.}
\end{figure*}

\begin{figure*}
    \centering
    \includegraphics[scale=0.55]{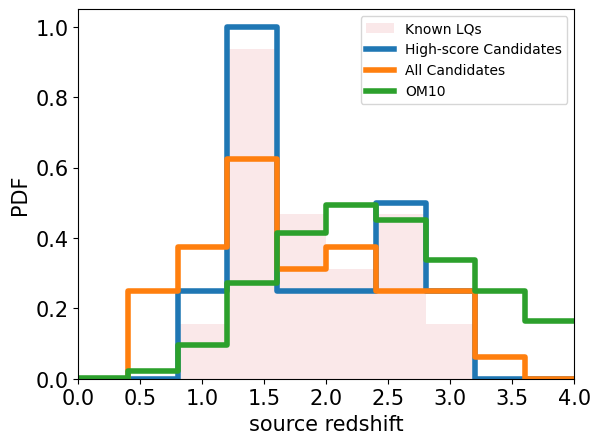}
    \includegraphics[scale=0.55]{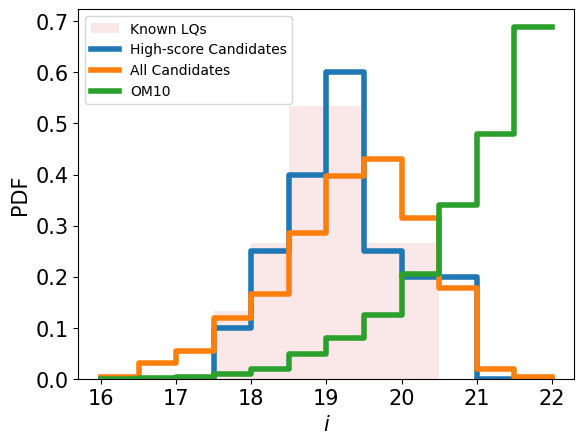}
    \includegraphics[scale=0.55]{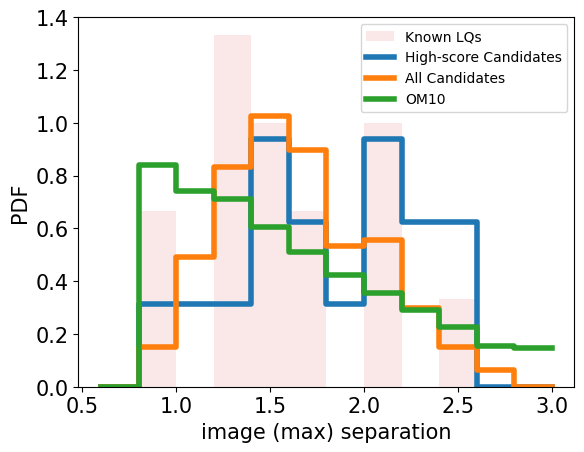}
    \includegraphics[scale=0.55]{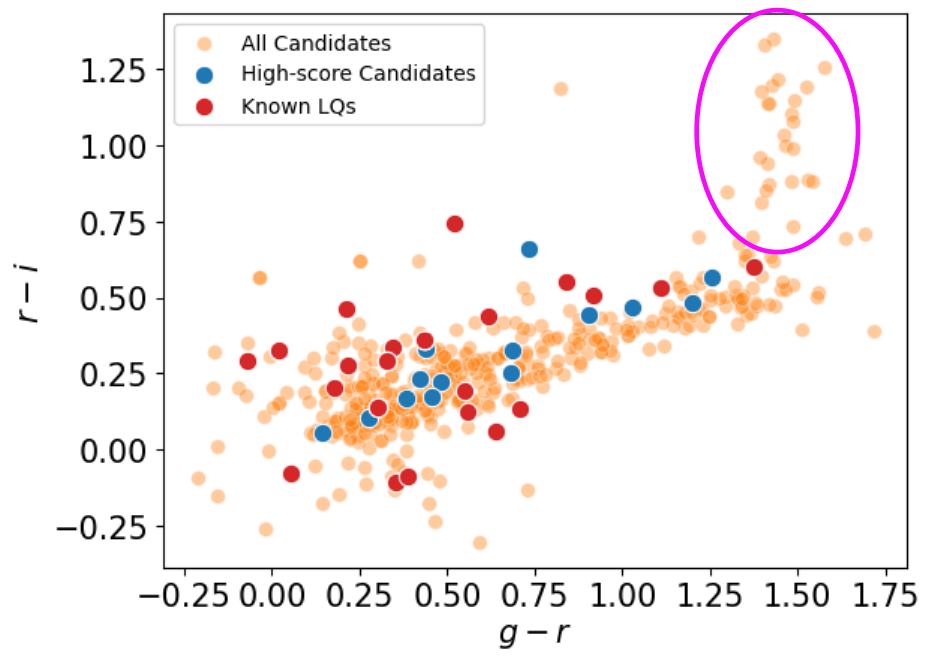}
    \caption{The distribution of $z_s$, $i$-band magnitude, image (maximum) separation, and $g-r$ vs. $r-i$ colors for our candidates is compared with known lensed quasars in the test set (see details in Table \ref{tab:real_lensed quasars}). The theoretical predictions from \cite{Oguri2010} is also plotted, as denoted as OM10 in the legend. In this plot, we focus on the color and magnitude of the quasars, excluding the magnitude and color of the lensing galaxies.}
    \label{fig:cand_pro}
\end{figure*}

\section{The Selection Effect of CNN}
\label{sec:sel_eff}

We evaluated the Network's selection effects of four parameters (image separation, flux ratio between the first and second brightest image, magnitude of the second brightest image, and flux ratio between the brightest image and the lensing LRG) using the lenses in simulated test set. Every tested parameter was divided into four or five bins, ensuring each bins have at least 100 simulated lensed quasars. 

\begin{figure*}
    \includegraphics[scale=0.55]{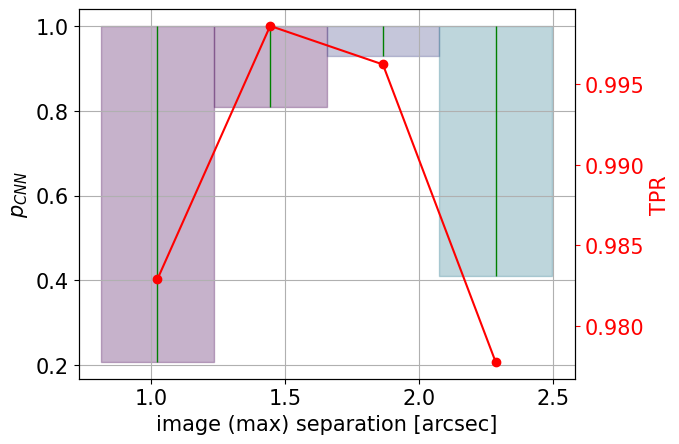}
    \includegraphics[scale=0.55]{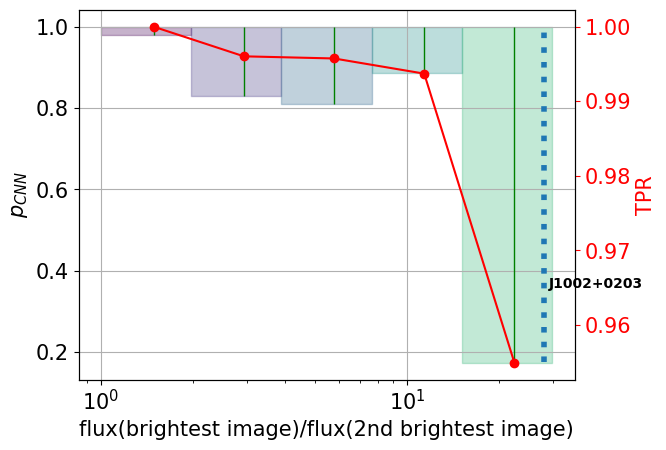}
    \includegraphics[scale=0.55]{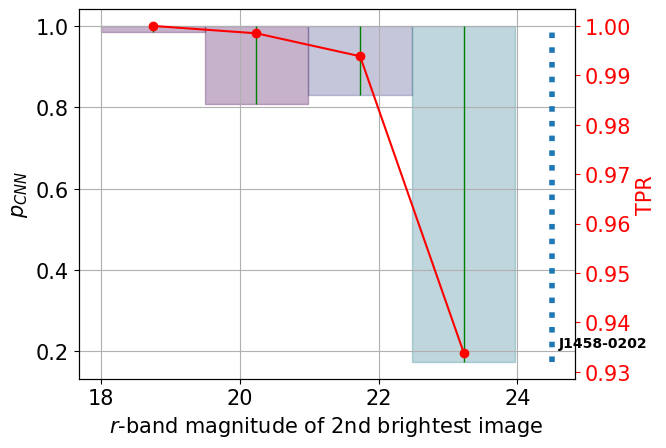}
    \includegraphics[scale=0.55]{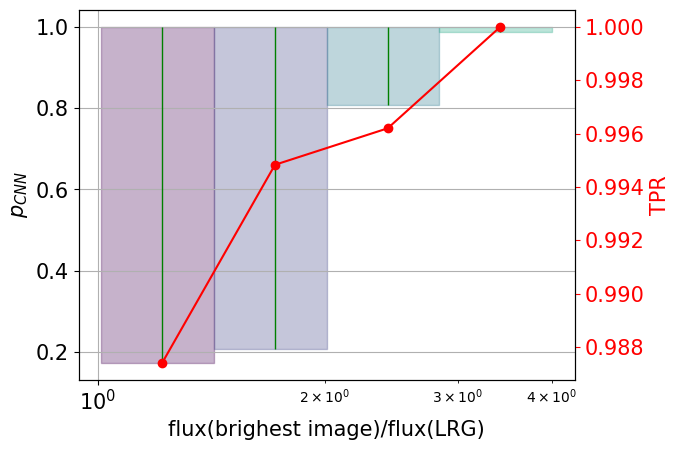}
    \caption{The $p_{CNN}$ and TPR in different parameter spaces is depicted. The green lines represent the range of $p_{CNN}$ in different bins, with the color beneath indicating various bin ranges. Each of the bin contains at least 100 data-points. The TPR is indicated by the red lines, with corresponding numerical values also displayed in red on the right $y$-axis. The upper left plot corresponds to image separation, followed by (upper right) the flux ratio between the brightest image and the second brightest image. The bottom left plot represents the magnitude of the second brightest image, while the bottom right one illustrates the flux ratio between the second brightest image and the lensing galaxy.}
    \label{fig:selection_effect}
\end{figure*}

The shaded areas with different colors in Figure \ref{fig:selection_effect} illustrate the range of different bins. On the left $y$-axis, the green vertical lines indicate the range (maximum and minimum values) of $p_{CNN}$, which is the probability given by the Network, versus the tested parameters. The TPRs calculated in different bins are depicted using a red line, with values shown on the right $y$-axis, also in red. We evaluated the variation of TPR due to changes in a tested parameter by comparing the highest and lowest TPRs across all tested bins:

\begin{equation}
    V_{TPR}=\frac{max(TPRs)-min(TRPs)}{max(TPRs)}.
\end{equation}

Regarding image separation, the Network performs best in the separation bin ($1.3\arcsec, 1.7\arcsec$). This aligns with the training set, which has the most samples near this separation (see the top right panel of Figure \ref{fig:properties_of_mock_lensed quasars}). The worst performance occurs in the ($2.1\arcsec, 2.5\arcsec$) bin, resulting in $V_{TPR}=2.2\%$.

For the flux ratio between the first and second brightest images, the TPR generally decreases as the flux ratio increases, with a sharp decline observed when the flux ratio exceeds 10. The $V_{TPR}$ for this parameter is 4.5\%. Regarding the magnitude of the second brightest image, the TPR generally decreases as the magnitude becomes fainter, with a sharp decline observed when the magnitude is fainter than 22. The $V_{TPR}$ for this parameter is 6.4\%. For the flux ratio between the brightest image and the LRG, performance improves as the flux ratio increases, with $V_{TPR}$ approximately 1.3\%.

Overall, performance varies across all tested parameters. The magnitude of the second brightest image has the largest $V_{TPR}$ ($\sim 6\%$), followed by the flux ratio between the first and second brightest images ($\sim 4\%$). Max separation and the flux ratio between brightest image and lensing LRG shows $V_{TPR}$ of 2.2\% and 1.3\% respectively.

Although the real test set is small in volume, it is worth examining why certain systems are missed by the Network. Three such systems (J1002+0203, J1458-0202, J2344-3056) are highlighted in Figure \ref{fig:real_lensed quasars} with yellow boxes only. Details of these systems can be found in Table \ref{tab:real_lensed quasars} and their corresponding references.

J1002+0203 is likely missed because its flux ratio is too large, leading to confusion in the Network. J1458-0202 was missed due to the very faint magnitude of one of its multiple images (approximately 24.5 in the $r$-band). J2344-3056, however, was missed due to its small image separation. The $\theta_E$ of this system is approximately $0.5$ arcsec, which falls below the training sample's selection criterion for $\theta_E$.

It is worth noting that our Network can recover three lensed galaxies (J2329-3409, J1224+0500, J1244+0106) listed in Table \ref{tab:real_lensed quasars}, even though the training set does not include such types of samples. This suggests that the Network possesses a degree of extrapolation capability.

\section{Summary}
\label{sec:sum}

We developed a ResNet-18 classifier \citep{Li2020} to identify strongly lensed quasar candidates in 3-band ($gri$) combined images from the KiDS Data Release 5. Simulated lensed quasars were generated using a mock catalog of lensed quasars with parameters reflecting realistic distributions. Based on the catalog, the images were created by combining real images of LRGs and simulated noisy multiply-imaged quasars. Negative samples in the training and validation sets were collected from real non-lenses. Sixty percent of the negative and positive samples are used for training, 10\% for validation, 30\% for constructing a simulated test set. Beside, a real test set that made by spectroscopy confirmed lensed quasars and non-lenses from KiDS observations is also prepared to better understanding the performance and fine-tuning the Network.

After the training and fine-tuning process, the Network was applied to target data, identifying 4,229 candidates. These candidates were visually inspected by first author Z.H. to eliminate the most unlikely ones. The remaining 3,172 objects were then evaluated by 10 inspectors, ultimately resulting in the selection of 272 candidates.
The $i$-band magnitude distribution of the lensed quasar candidates predominantly ranges from 18 to 21, peaking at approximately 20. The maximum separation between images varies from 1.0 to 2.5$''$, with a peak at 1.5$''$. The redshift distribution of quasars, assessed from a subset of samples with spectroscopic redshifts (comprising 40 systems), spans from 0.422 to 3.57, with a peak near 1.5.
According to \cite{Oguri2010}, approximately 170 lensed quasars remain to be discovered within the KiDS DR5 footprints. After excluding known systems previously identified in earlier studies from the initial pool of 272 candidates, the total is reduced to 229. This reduction will help fill the pool of undiscovered quasars, particularly for $i$ magnitudes exceeding 20.5.

The Network demonstrated strong performance in rediscovering known systems, as evaluated by a real test set. It successfully rediscovered 81.25\% of known systems in the test set. Simultaneously, the FPR was only approximately 1/2000.
Furthermore, a cross-match with SLED revealed the following:
Among the high-score candidates, there was one overlap and no contaminants.
Among the median-score candidates, there were 15 overlaps, with 7 out of 15 being contaminants.
Among the lower-score candidates, there were 11 overlaps, with 8 out of 11 being contaminants.
This indicates that lower-score candidates are more likely to be polluted non-lenses compared to median-score ones. More precise purity assessments will be possible once spectroscopic follow-ups are conducted.

Through testing and analysis, we observed stable performance of our classifier across four parameters (see the details at Section \ref{sec:sel_eff}). The largest degradation ($\sim 6\%$) of TPR is observed  when the magnitude of second brightest quasars exceed $r\sim 23$. The flux ratio between two images also has a influence: when exceed $\sim 10$, the TPR will decrease to $\sim 0.95$ from 1, corresponding to a $\sim 5\%$ degradation. The image separation and the flux ratio between LRG and brightest quasar image, have $\sim 2\%$ influence. The selection bias of image separation has also been observed in other works \citep[e.g.,][]{Herle2024}. Furthermore, analysis of real test sets revealed identifiable reasons for the classifier's omission of three spectroscopically confirmed lensed quasars, attributed to factors such as large flux ratio between images, faintness of the second brightest quasar image, or small image separations. Notably, our classifier demonstrated the ability to rediscover three lensed compact galaxies, despite their exclusion from the training data, showcasing its robustness and adaptability.

The lensed-quasar candidates discovered in this paper will be analyzed in more details in the future. Firstly, image decomposition will be performed to check whether any lensing galaxies can be detected.. Secondly, the color and astrometric information will be analyzed to further eliminating star contaminants. Based on the results of image decomposition, color and astrometric analysis, the follow-up spectroscopy will be conducted to confirm the most promising candidates. In this work, we focused on the discovery of the systems with $\theta_E$ larger than $0.6''$, which is expected to have clear multiple images in KiDS survey. As one of our future plans, the methodology will also be improved to search for smaller image separation systems (for instance, maximum separation larger than $0.6''$) by constructing targeted training and testing samples. Those lensed quasars are harder to identify, but still possible.

In summary, we constructed a CNN classifier to identify strongly lensed quasar candidates in KiDS DR5 imaging data, resulting in the discovery of 229 new candidates. The training set preparation pipelines and fine-tuning methodologies developed in this work can be applied to further lensed quasar identification efforts in other wide-field imaging surveys, such as Euclid \citep{Euclid-intro,Euclid2024}, CSST, Wide-Field Survey Telescope \citep[WFST,][]{WFST-intro}, and Dark Energy Spectroscopic Instrument Legacy Imaging Surveys \citep[DESI-LS,][]{Dey2019}. It has been predicted that numerous strong lenses will be identified from those surveys \citep{Collett2015,Cao2024}.
Future large-scale spectroscopic surveys, such as DESI and the 4-meter Multi-Object Spectrograph telescope \citep{deJong2019}, will help confirm a portion of these candidates. Additionally, targeted observations using telescopes like P200 and New Technology Telescope will also assist in confirming some candidates. The confirmed systems from this work can potentially be used to study the circumgalactic medium \citep{Cai2019,Lau2022}, the accretion disks of background AGN \citep{Sluse2017,Fian2021}, and the Hubble constant \citep{Suyu2017,Liao2019,Wong2020}.

\section*{Acknowledgements}

We thank the referee in the KiDS internal
review process for detailed comments, which
helped improve the quality of this paper. We gratefully acknowledge Dr. C. Lemon for assisting us in cross-matching our candidate list with the SLED database, which we have found immensely helpful for our study. Additionally, Dr. Lemon provided valuable suggestions regarding some of the candidates. We also extend our thanks to Dongxu Zhang, Ran Li, Xiaoyue Cao, and Zuhui Fan for their constructive discussions.\\

Z.H. acknowledges that this work is supported by the Postdoctoral Fellowship Program of CPSF under Grant Number GZC20232990. R.L. acknowledges the support of the National Nature Science Foundation of China (No 12203050). Y.S. acknowledges the support from the China Manned Space Program through its Space Application System. S.S. has received funding from the European Union’s Horizon 2022 research and innovation program under the Marie Skłodowska-Curie grant agreement No 101105167 - FASTIDIoUS. C.T. acknowledges the INAF grant 2022 LEMON. G.L. acknowledges the support of the National Nature Science Foundation of China Manned Spaced Project (CMS-CSST-2021-A12). N.L. acknowledges the support of the CAS Project for Young Scientists in Basic Research (No. YSBR-062), the science research grants from the China Manned Space Project (No. CMS-CSST-2021-A01), and the hospitality of the International Centre of Supernovae (ICESUN), Yunnan Key Laboratory at Yunnan Observatories, Chinese Academy of Sciences. A.D. acknowledges the Grant of ERC Consolidator (No.770935).


\bibliography{cite-database}
\bibliographystyle{aasjournal}


\appendix 
\section{One quad candidate}
\label{sec:app}

The quad candidate at R.A. = 12.626$^\circ$ and dec. = -30.630$^\circ$ is detected as a single source in TILE KIDS\_12.7\_-30.2, with magnitudes of $u=20.798\pm 0.017$, $g=20.025\pm 0.003$, $r=19.698\pm 0.002$, $i=19.570\pm 0.008$, $z=19.511\pm 0.010$, $Y=19.432\pm 0.020$, $J=19.337\pm 0.017$, $H=19.340\pm 0.027$, $Ks=19.198\pm 0.044$.

\begin{figure}
    \centering
    \includegraphics[width=0.5\linewidth]{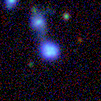}
    \caption{The quad candidate in the southern sky was accidentally discovered during the network fine-tuning process. The image size is 20\arcsec by 20\arcsec, and its orientation and plotting scheme are consistent with the ones that were used in Figure \ref{fig:real_lensed quasars}.}
    \label{fig:enter-label}
\end{figure}

\end{document}